\documentclass[11pt]{article}
\usepackage{graphicx}
\usepackage{epsf}
\usepackage{epsfig}

\newcommand{\ds}{\displaystyle }
\newcommand{\R}{{\sf R\hspace*{-0.94ex}%
\rule{0.15ex}{1.5ex}\hspace*{0.94ex}}}
\newcommand{\Z}{{\sf Z\hspace*{-0.94ex}%
\rule{0.15ex}{1.5ex}\hspace*{0.99ex}}}
\newcommand{\N}{{\sf N\hspace*{-0.99ex}%
\rule{0.15ex}{1.5ex}\hspace*{0.99ex}}}

\title{A regularized representation of the fractional Laplacian in $n$ dimensions and its relation to Weierstrass-Mandelbrot type fractal functions}

\author{ {\sl Thomas M.  Michelitsch$^{1,2}$\footnote{Corresponding author, e-mail~: michel@lmm.jussieu.fr, www~: http://bit.ly/champs-fractelysees}, G\'erard A. Maugin$^{1,2}$, Shahram Derogar$^{3}$, Mujibur Rahman$^{4}$, } \\ \\
$^1$ Universit\'{e} Pierre et Marie Curie, Paris 6\\
$^2$Institut Jean le Rond d'Alembert, CNRS UMR 7190 \\
4 Place Jussieu \\
75252 Paris cedex 05 \\
 FRANCE\\ \\ \\
$^{3}$
Department of Architecture\\
Yeditepe University\\
26 August Campus \\
34755 Kayisdagi - Atasehir\\
Istanbul \\
TURKEY
\\ \\ \\
$^4$
General Electrics Energy\\
300 Garlington Road \\
 Greenville, SC 29615 \\
            USA
\\ \\ {\it IMA Journal of Applied Mathematics
(2014) 79, 753-777.  \, doi:10.1093/imamat/hxu018 } 
}
\begin{document}
\maketitle

\paragraph{Abstract}
We demonstrate that the fractional Laplacian (FL) is the principal characteristic operator of harmonic systems with {\it self-similar}
interparticle interactions. We show that the FL represents the ``{\it fractional continuum limit}'' of a discrete ``self-similar Laplacian" which is obtained by Hamilton's variational principle from a discrete spring model.
We deduce from generalized self-similar elastic potentials regular representations for the FL which involve convolutions of symmetric finite difference operators of even orders extending the standard representation of the FL.
Further we deduce a regularized representation for the FL $-(-\Delta)^{\frac{\alpha}{2}}$ holding for $\alpha\in \R \geq 0$.
We give an explicit proof that the regularized representation of the FL gives for integer powers $\frac{\alpha}{2} \in \N_0$ a distributional representation of the standard Laplacian operator $\Delta$ 
including the trivial unity operator for $\alpha\rightarrow 0$. 
We demonstrate that self-similar {\it harmonic} systems are {\it all} governed in a distributional sense by this {\it
regularized representation of the FL} which therefore can be conceived as characteristic footprint of self-similarity.

\paragraph{Keywords:}
\noindent {\it  Fractional Calculus, Fractional Laplacian, Regularization, Weierstrass-Mandelbrot fractal functions, Self-similarity, Distributions, 
Generalized functions, L\'evy flights}.

\section{Introduction}

Fractional calculus has a long history, but it took until the last decade for a new interest to emerge to employ fractional operators and the so called {\it fractional Laplacian}
(FL) (often also referred to as Riesz fractional derivative) $-(-\Delta)^{\frac{\alpha}{2}}$ where $\frac{\alpha}{2} \in \R$ indicates a fractional, in general non-integer, power of the standard Laplacian $\Delta$ where the sign is 
chosen such that
the FL is for $\alpha>0$ a negative (semi-) definite self-adjoint operator, formally defined by its Fourier representation $-k^{\alpha} \leq 0$.
The {\it positive} exponent $\alpha >0$ is a requirement which comes into play when translational invariance (uniform translations corresponding to the $k=0$ mode)
give a a vanishing contribution.
The Green's function kernel of its inverse operator $-(-\Delta)^{-\frac{\alpha}{2}}$, the so-called Riesz potential, is also well known in the literature and was first
introduced by the Hungarian mathematician Marcel Riesz \cite{riesz}. 

This newly emerging interest in fractional calculus is due to the fact that it has been recognized that critical and complex phenomena
with self-similarity properties can be often described by fractional equations of motion; especially phenomena of anomalous diffusion (L\'evy flights) are governed
by fractional diffusion equations.

One incentive to ``popularize'' fractional calculus in a broader physics and engineering community was probably the survey article of Metzler and Klafter \cite{metzler} with a broad overview on applications demonstrating the power of the fractional approach to model complex phenomena in various fields of physics and especially the interlink between self-similar processes such as for instance phenomena of anomalous diffusion (L\'evy flights) and its fractional continuum description.
Prominent examples of L\'evy flights are phenomena as various as the time evolution of the stock market \cite{mandel} and the chaotic particle transport in turbulent flows \cite{chen-fract-turb}.
Recently many models where developed which employ the {\it FL} in various physical contexts, among them the description of ``complex" dynamic phenomena \cite{chen,chen2,chen3,hanyga,brockmann,stollen,vazquez,riesz2,reichel,lu,michel-fcaa} and see also the numerous references therein.
We also would like to mention the recently emerging great impact of {\it fractional Quantum Mechanics} coined by Laskin \cite{laskin}. 

In a recent article we exhibited the interlink
between harmonic self-similar spring systems with elastic potentials of Weierstrass-Mandelbrot function type and the FL defined rigorously by the ``fractional continuum limit'' \cite{michel-fcaa} which will be the starting point of the present paper.
Here our aim is to demonstrate that the FL is a natural consequence of self-similarity (scale-freeness) of elastically harmonic systems. We deduce a regularized representation of the FL which holds for any positive exponent $\alpha >0$ extending the standard representation of the FL which is limited
to exponents $0<\alpha<2$.

So far, the spatial representations of the FL are limited due to its hypersingular characteristics to $0<\alpha<2$ where $\alpha\rightarrow 2$ approaches asymptotically the standard Laplacian $\Delta$ and $\alpha\rightarrow 0$ the unity operator. For a mathematical analysis on some key properties of the FL we refer to textbooks such as \cite{sylvestre}.
Generally the FL is a nonlocal convolutional and highly singular operator. Its standard spatial representation given in the literature exists in its non-regularized form only in the interval $0<\alpha<2$\footnote{In the stochastic context of L\'evy statistics the range $0<\alpha<2$ is refered to as ``L\'evy interval" \cite{mandel}.}, e.g. \cite{vazquez}. 
In many {\it physical contexts} power law spectra and power law distributions indeed are found within this range \cite{clauset}.  However, examples exist which justify to analyze the FL also outside the $\alpha$-range $(0,2)$: Gopikrishnan et al. found a power law distribution of fluctuations of financial market indices which exhibit a scaling exponent $\alpha$ being well {\it outside} the L\'evy range $0<\alpha<2$ \cite{gopikrishnan}. These probablity distributions are described by characteristic functions (Fourier transform of the probability distributions) given by exponentials $\sim \exp{-k^{\alpha} \tau}$ (with the scaled time $\tau \geq 0$) with scaling exponents $\alpha$ outside the L\'evy r\'egime $(0,2)$. Such stochastic processes have to be described by a Fokker-Planck equation which employs a spatial representation for the FL which exists {\it outside $0<\alpha<2$}. 
To render such problems accessible to the analysis, it is necessary to have such a spatial
representation for the FL with an extended $\alpha$-range of validity, and it is more than ever desirable to have a spatial representation of the FL which covers the entire range of principally admissible exponents $\alpha >0$. Such a representation for the FL outside the L\'evy range $(0,2)$, however, requires regularization since the standard representation of the FL, in its common unregularized form,  does not exist outside of this interval $0<\alpha<2$ ({\it more precisely: for $\alpha >2$ the convolutional integral of the unregularized standard representation of the FL diverges as $\sim r^{2-\alpha}$ at $r\rightarrow 0$}). The goal of the present paper is hence to deduce by means of a simple regularization rule a regularized unique {\it spatial representation} for the FL which holds for
any $\alpha >0$.

In the literature as well as in the present paper we define the FL such that it has all ``good properties`` of the elliptic, 
standard Laplacian (having Fourier transform $-k^2$). These good properties of the FL are: self-adjointness, spatial isotropy and negative-semidefiniteness 
(``semi-" since the Fourier transform $-k^{\alpha}$ of the FL is vanishing at $k=0$ reflecting translational invariance). The latter property requires positive exponents $\alpha >0$
\footnote{which are in a physical discrete spring model picture equivalent to a self-similar 
distribution of interparticle springs decaying with interparticle distance. Allthough $\alpha=0$ is due to translational invariance physically a forbidden value, 
the regularized representation of the FL deduced in this paper holds for all $\alpha \geq 0$, (including $\alpha=0$).}.

In the present paper the {\it FL operator} is formally defined as the operator $-(-\Delta)^{\frac{\alpha}{2}}$ having negative eigenvalues $-k^{\alpha}=-(k^2)^{\frac{\alpha}{2}} \leq 0$ for any $\alpha \in \R >0$ (vanishing only for $k=0$ when $\alpha>0$),
leading to all above mentioned good properties of a Laplacian. However, this formal definition of the FL does not provide any spatial representation of this operator.
Our main goal is to deduce spatial representations for the FL from simple elastic potentials in a rigorous manner by application of Hamilton's variational principle.
We will show in the present paper that the
FL is the ``natural" operator, and in a sense the most basic operator that is 
generated as a continuum limit from a physical ``self-similar" spring model and its generalizations.
\\ \\
The paper is organized as follows: As point of departure in section \ref{linchain} we evoke a 1D harmonic spring model with harmonic elastic potential energy which describes {\it self-similar interparticle interactions} 
which we developed recently \cite{michel}. This discrete model leads to fractal dynamic vibrational characteristics such as a dispersion relation of the form of Weierstrass-Mandelbrot fractal functions. 
Then we analyze a ``fractional continuum limit" which we introduced earlier \cite{michel-fcaa} where the self-similar Laplacian of the discrete model takes up to a normalization factor the form of the standard 
representation of the FL which is employed in many references, e.g. \cite{chen,hanyga,brockmann,stollen,vazquez,michel-fcaa}
with restricted exponent $0<\alpha<2$. We demonstrate that the discrete fractal model is the natural discrete counterpart of the fractional model. 
Both models are linked by the ``fractional continuum limit" which we introduced recently \cite{michel-fcaa}.
Further in section \ref{gen1}, we deduce spatial representations of the FL which hold for an extended range of exponents $\alpha$. To this end we introduce generalized discrete self-similar harmonic potentials 
which involve higher-order finite differences in the displacement field of integer orders $m \in \N$ and analyze the fractional continuum limit: 
Hamilton's variational principle then defines in a rigorous manner self-similar Laplacians leading in the fractional continuum limit to new representations for the FL which exist for exponents
$0<\alpha<2m$ where $m\in \N$ indicates arbitrary integers. These representations depend on the order $m\in \N$ of the finite difference operator involved, where $m=1$  coincides with the standard representation of the FL  \cite{michel,michela,michelb,michelc,michel-fcaa}. 
We show by means of a regularisation rule that these new representations of the FL are all equivalent to a single {\it unique regularized representation of the FL} holding for $\alpha \geq 0$. We refer to this regularized representation of the operator $-(-\Delta)^{\frac{\alpha}{2}}$ as Fractional Laplacian (FL) despite it holds for both, fractional and integer powers of the standard Laplacian.

The regularization elaborated in the present analysis concerns even (isotropic) functions with respect to the hyper-singular point. This regularization makes sense for ``functions under the integral'', i.e. we deal with {\it distributions}
(generalized functions) in the generalized sense as presented by G'elfand and Shilov \cite{gelfand}. The regularization concerns tempered distributions being defined in the {\it Schwartz space} 
(space of functions which have a Fourier transform, see e.g. \cite{rudin}).

\subsection{Standard representation of Fractional Laplacian}

First of all let us evoke the standard representation of the {\it FL} $-(-\Delta)^{\frac{\alpha}{2}}$ of the multidimensional space $\R^n$ defined for sufficiently good functions $\Phi$ as (e.g. \cite{chen,chen2,chen3,hanyga,vazquez,reichel,michel-fcaa})

\begin{equation}
 \label{sing1}
-(-\Delta)^{\frac{\alpha}{2}} \Phi({\bf x}) = \frac{{\cal C}_{(n,\alpha)}}{2}\int \frac{\Phi({\bf x}+{\bf r})+
\Phi({\bf x}-{\bf r})-2\Phi({\bf x})}{r^{n+\alpha}}{\rm d}^n{\bf r}       ,\hspace{2cm} 0<\alpha<2, n\in \N
\end{equation}
where ${\bf r}\cdot{\bf r}=r^2$ and ${\cal C}_{(n,\alpha)} >0$ is a positive normalisation factor. With ``sufficiently good functions" we mean smooth fields $\Phi$
with the properties
\begin{equation}
\label{addprop}
\begin{array}{l}
\displaystyle \lim_{r\rightarrow 0}|\Phi({\bf x}+{\bf r})+
\Phi({\bf x}-{\bf r})-2\Phi({\bf x})| \sim r^2 \rightarrow 0 \nonumber \\ \nonumber \\
\displaystyle \lim_{r\rightarrow \infty} |\phi({\bf r})| = 0  
\end{array}
\end{equation}
This relation means that a sufficiently good field $\Phi$ fulfills subsequently introduced general conditions (\ref{asymptotics}) (where $a=0<\alpha<b=2$) defining the admissible function space in which
the standard representation of the FL (\ref{sing1}) is well defined.
We observe that with (\ref{addprop}) the $r$-integration of (\ref{sing1}) remains finite at infinity ($r=|{\bf r}|$), so that 
\begin{equation}
\label{intex}
|\int_0^{const} \frac{r^2 r^{n-1}}{r^ {\alpha+n}}{\rm d}r |\sim r^{2-\alpha}|_{r=0}^{const} < \infty \, {\rm if}\, , \alpha <2
\end{equation}
where we plugged in (\ref{addprop})$_1$, 
and at infinity the $r$-integration of (\ref{sing1}) behaves as
\begin{equation}
\label{outex}
|\int_{const}^{\infty} \frac{\Phi({\bf x}+{\bf r})+
\Phi({\bf x}-{\bf r})-2\Phi({\bf x})}{r^{\alpha+1}}{\rm d}r| \sim |-2\phi({\bf x}) r^{-\alpha}|^{\infty}_{const}| < \infty \, 
{\rm if}\,\, \alpha >0 
\end{equation}
where we plugged in (\ref{addprop})$_2$.
It follows from (\ref{intex}), (\ref{outex}) that the standard FL (\ref{sing1}) exists in the interval $0<\alpha<2$ for sufficiently regular fields $\Phi$, i.e. which fulfill (\ref{addprop}).

We follow in this paper the convention that the FL is formally defined as operator with Fourier representation $-k^{\alpha}=-(k^2)^{\frac{\alpha}{2}} \leq 0$ ($k=|{\bf k}|$).
The sign convention of the FL is here defined in such a way that it has all good properties of a Laplacian\footnote{In some references the FL is defined by an opposite sign.}. We come back to the crucial properties of the FL
in the course of this paper.
The normalization factor is obtained as, e.g.  \cite{vazquez,michel-fcaa}

\begin{equation}
 \label{normfac}
{\cal C}_{(n,\alpha)} =  \frac{2^{\alpha-1}\alpha\Gamma(\frac{\alpha+n}{2})}{\pi^{\frac{n}{2}}\Gamma(1-\frac{\alpha}{2})} > 0 ,\hspace{2cm} 0<\alpha<2,
\end{equation}
where the $\Gamma$-function 
is defined by \cite{abramo}

\begin{equation}
 \label{gamma}
\alpha! =\Gamma(\alpha+1)=\int_0^{\infty}\tau^{\alpha}e^{-\tau}{\rm d}\tau ,\hspace{2cm} \alpha >-1
\end{equation}
which exists for $\alpha >-1$ and takes for $\alpha\in \N_0$ the values of the usual integer factorial with $0!=1$ which follows from its recursive behavior $\alpha!=\alpha(\alpha-1)!$.

The integration in (\ref{sing1}) is performed over the entire multidimensional space $\R^n$.
Another useful equivalent representation of the FL is

\begin{equation}
 \label{singular}
-(-\Delta)^{\frac{\alpha}{2}} \Phi({\bf x}) =: {\cal C}_{(n,\alpha)}\int \frac{\Phi({\bf r})-\Phi({\bf x})}{|{\bf r}-{\bf x}|^{n+\alpha}}{\rm d}^n{\bf r}       ,\hspace{2cm} 0<\alpha<2, n\in \N
\end{equation}
where $n\in \N$ denotes the dimension of the physical space. In this paper we refer to (\ref{sing1}) and (\ref{singular}) as the {\it standard representation of the FL} which exists for sufficiently good functions for $0<\alpha<2$.

\section{Deduction of the FL from a self-similar linear spring model}
\label{linchain}

In order to motivate the point of departure physically, we briefly evoke some principal aspects of the self-similar spring model introduced in \cite{michel}.
To this end consider a spatially homogeneous distribution of mass with constant density
$\rho_0$. Each mass-point is represented by the spatial coordinate $x$
and is only connected with mass points of distances $l_s=h a^s$ ($h,a>0$) located at $x\pm h a^s$ where $s\in \Z_0$ passes all positive and negative integers. 
We note that by construction the entire set of connecting distances $\{l_s\}$ ($s\in \Z_0$) is
invariant under the rescalings $\{l_s\} \rightarrow \{a^pl_s\}$ with $p\in \Z_0$ being integer powers of a prescribed scaling factor $a\in\R$. Note that $a$ and $a^{-1}$ define the same set of sequences $\{l_s\}$ since only $s\rightarrow -s$ is exchanged.
The harmonic springs connecting the material point $x$ with the material points $x\pm h a^s$
are assumed to have the spring constants $f_0 a^{-\delta s}= f_0(l_s/h)^{-\delta}$, i.e. scaling with a power law with respect to the distances $l_s$. It appears necessary that only $\delta>0$ is {\it physically admissible}, as the spring constants should decay with increasing particle distance $l_s$ and vanish for $l_s\rightarrow \infty$ ($s\rightarrow \infty$).

Let $u(x)$ denote the displacement field associated with material point $x$ (or in a generalized context, the field variable associated with $x$). 
Due to the continuous spatial mass distribution the described harmonic spring system can be conceived as a hybrid between lattice and continuum: 
With respect to the spatial physical properties it is a continuum, i.e. all physical quantities depend on the continuous variable $x$, however, as we will see, 
the distribution of springs versus distance is a discrete {\it self-similar distribution}.
The elastic energy density associated to mass point $x$ is proposed in the form of a Weierstrass-Mandelbrot type function \cite{michel}

\begin{equation}
 \label{Welast}
\ds {\cal W}(x,h)=\frac{f_0}{4}\sum_{s=-\infty}^{\infty} a^{-\delta s}\left\{(u(x+ha^s)-u(x))^2+(u(x-ha^s)-u(x))^2\right\}
\end{equation}
which converges for sufficiently good fields $u(x)$ in the interval $0<\delta<2$.
The total elastic energy is given by
\begin{equation}
\label{elastenergytot}
 V(h)= \int_{-\infty}^{\infty} {\cal W}(x,h)\,{\rm d}x
\end{equation}

Let us put here for our convenience the mass density $\rho_0=1$ and spring constant $f_0=1$ being equal to unity. The prefactor of $1/4$ takes into account the double counting of interparticle springs. (\ref{Welast}) is invariant by replacing $a\rightarrow a^{-1}$, therefore we can confine without loss of generality to $a\in \R >1 $.
The elastic energy density (\ref{Welast}) is constructed such that it constitutes a {\it self-similar function} with respect to $h$, and fulfills the following self-similarity condition \cite{michel}

\begin{equation}
\label{self-sim}
\ds {\cal W}(x,ah)= a^{\delta} {\cal W}(x,h)
\end{equation}
which implies validity of this relation for all integer powers $a^s$ ($s\in \Z_0$). Self-similarity of a function $\Lambda(ah)= a^{\delta}\Lambda(h)$ is equivalent with {\it Log-periodicity}
($\ln(a)$-periodicity) of the function $g(\ln(h))=\ln(h^{-\delta}\Lambda(h)) =g(\ln(h)+\ln(a))$ with respect to $\ln(h)$. Log-periodicity is a typical footprint of discrete scale-invariances observed in turbulence.
An interesting speculative study on this subject and its link to the FL has been presented recently by Chen \cite{chen-fract-turb}.

A function $\Lambda(h)$ which is self-similar in the sense (\ref{self-sim}) is refered to as {\it self-similar at the point $h=0$}. The corresponding notion of self-similarity at a point probably was coined by Peitgen, J\"urgens and Saupe \cite{peitgen}.
We notice that the self-similar elastic energy density (\ref{Welast}) can be in some cases a fractal function with respect to $h$. The self-similarity condition (\ref{self-sim}) indicates the absence of a characteristic intrinsic interaction length scale. The variable $h$ is a dimensional parameter having dimension of length, however, without the meaning of a characteristic (intrinsic) length.

We define now a ''Laplacian" by application of Hamilton's variational principle on (\ref{elastenergytot}) for the field $u$ which leads to

\begin{equation}
\label{hamilton}
 \Delta_{(\delta,h)}u(x) =:-\frac{\delta V}{\delta u(x)}
\end{equation}
where $\frac{\delta}{\delta u}$ denotes the functional derivative with respect to the field $u$. We obtain in this way the {\it self-similar Laplacian}

\begin{equation}
 \label{laplacian1D}
\ds \Delta_{(\delta,a,h)}u(x) = \sum_{s=-\infty}^{\infty} a^{-\delta s}\left(u(x+ha^s)+u(x-ha^s)-2u(x)\right) ,\hspace{1cm} 0<\delta<2
\end{equation}
converging for the same sufficiently good fields $u(x)$ as (\ref{Welast}) in the interval $0<\delta<2$.
We refer to this operator as  {\it self-similar Laplacian} since it produces a self-similar function with the scaling property (self-similarity condition)

\begin{equation}
 \label{selsirelalap}
\Delta_{(\delta,a,ah)} = a^{\delta}\Delta_{(\delta,a,h)}
\end{equation}
holding only for a given $a$ and its (positive and negative) integer powers $a^s$ $s\in \Z_0$. This relation means 
nothing else but that changing of the scale
$h'=a^nh$ does not change, it just rescales the variables: The equation of motion is obtained from Hamilton's principle by relation (\ref{hamilton}) and takes the form
\begin{equation}
\label{autosimondes}
\frac{\partial^2}{\partial t^2}u(x,t)=\Delta_{(\delta,a,h)}u(x,t)
\end{equation}
We refer this equation to as {\it self-similar wave equation}. The dispersion relation is determined by $\Delta_{(\delta,a,h)}e^{ikx}=-\omega_{(\delta,a)}^2(k)e^{ikx}$ in the form \cite{michel}

\begin{equation}
 \label{dispersion}
\omega_{(\delta,a)}^2(kh)=4\sum_{s=-\infty}^{\infty}a^{-\delta s}\sin^2{(\frac{kha^s}{2})} ,\hspace{2cm} 0<\delta<2
\end{equation}

This dispersion relation is a Weierstrass-Mandelbrot function (examples are plotted in the figures) converging also in the range $0<\delta<2$. In the interval $0<\delta<1$ this function has a fractal graph of Hausdorff dimension $D=2-\delta>1$ which is nowhere differentiable \cite{hardy}. In the range $1<\delta<2$
(\ref{dispersion}) is a regular non-fractal (once differentiable) function. The fractal erratic behavior  of (15) appears only in the range $0<\delta<1$
and is caused by the critically slow decay of the powers $a^{-\delta s}$ as $s\rightarrow\infty$. For small exponents $\delta\rightarrow 0$ high frequency oscillations
are only little suppressed in the series (15) leading to more and more erratic features with
an increasing fractal Hausdorff dimension $D=2-\delta$ which approaches the plane filling dimension $2$. In its entire range of convergence $0<\delta<2$ the Weierstrass-Mandelbrot function type dispersion relation (\ref{dispersion}) 
is self-similar with respect to the non-dimensional wave number $kh$, namely $\omega_{(\delta,a)}^2(akh)= a^{\delta}\omega_{(\delta,a)}^2(kh)$.

In the two figures two dispersion curves are plotted, for fractal regime (erratic curve for first figure) $0<\delta<1$ and non-fractal regime $1\leq \delta<2$, respectively (erratic curve of second figure). The smaller $\delta$ the more erratic become
the dispersion curves.
The smooth curves indicate their fractional continuum limits $a\rightarrow 1$ (given in explicit form in below equation (\ref{scalingdis})) where the erratic fractal characteristics are wiped out.

\begin{figure*}[H]
\centering
\includegraphics[scale=1.0]{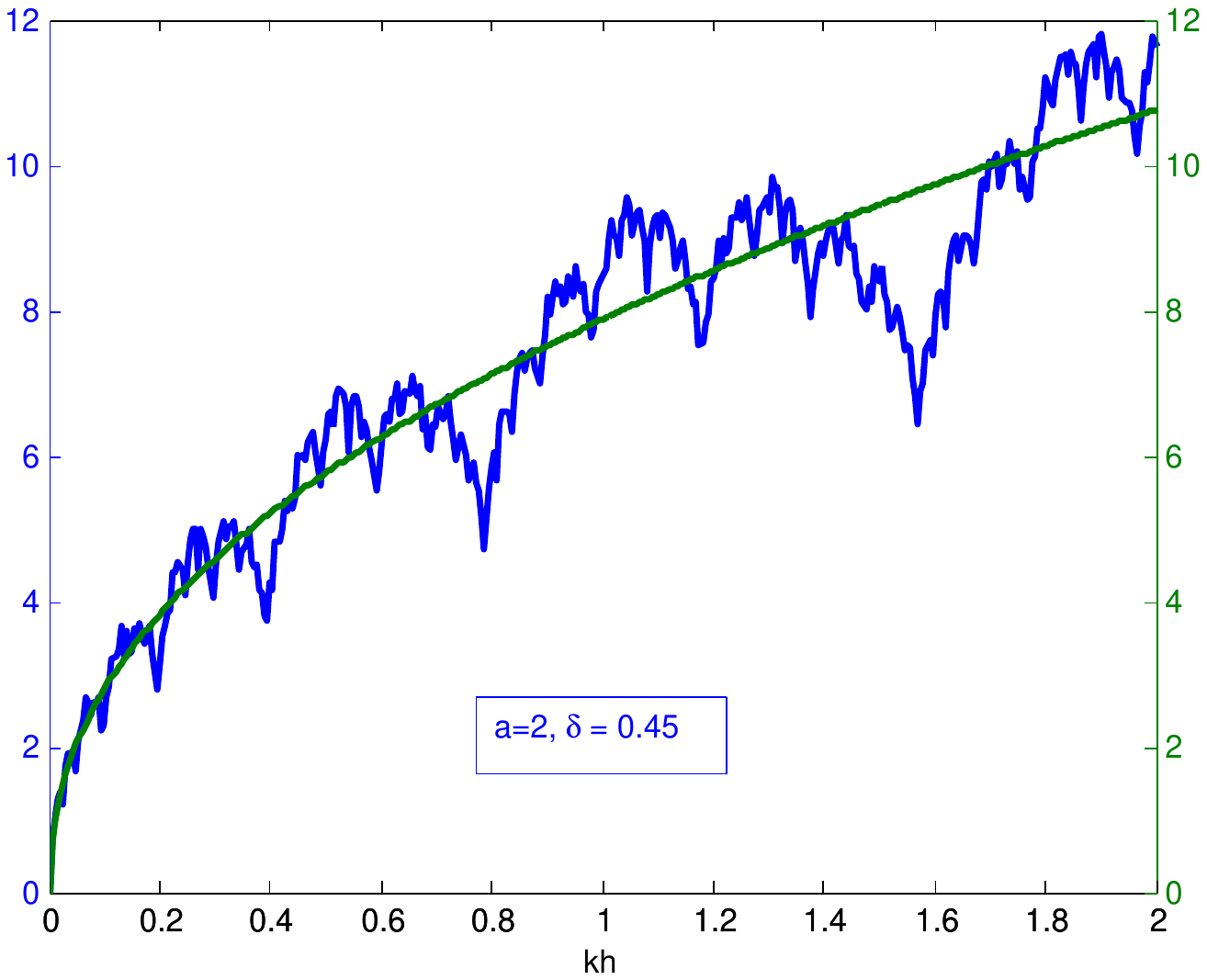}
\label{fig:1}
\end{figure*}
\begin{figure*}[H]
\centering
\includegraphics[scale=1.0]{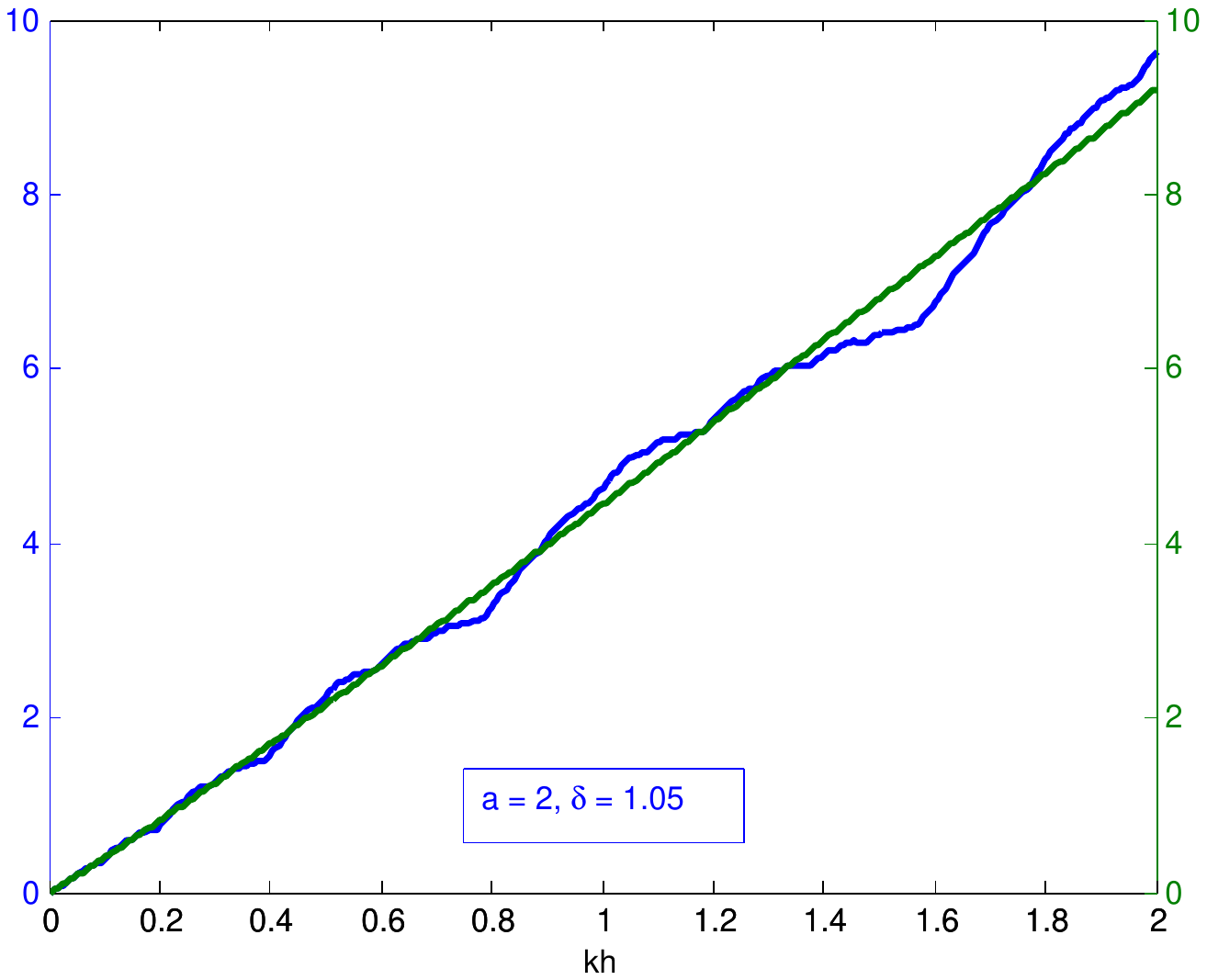}
\caption*{Dispersion relation of Weierstrass-Mandelbrot function type (\ref{dispersion}) for two cases: fractal regime $0<\delta=0.45<1$ (upper figure), nonfractal regime $1\leq \delta=1.05 <2$ (lower figure). 
The smooth curves indicated the corresponding power law dispersion relation $\omega_{(\delta,a)}^2(k)=A_{\delta}k^{\delta}$ of (\ref{scalingdis}) representing te fractional continuum limit of (\ref{dispersion}).}
\label{fig:2}
\end{figure*}

\subsection{Fractional continuum limit}
\label{fraclim}

An important limiting case which we refer to as the {\it fractional continuum limit} is obtained for $a\rightarrow 1$ in relations (\ref{Welast}), (\ref{laplacian1D}) and (\ref{dispersion}) \cite{michel-fcaa}.
This limiting case connects the self-similar discrete spring model of the subsequent section with fractional calculus,
especially the discrete 1D self-silmilar Laplacian (\ref{laplacian1D}) with its 1D FL counterpart.
To demonstrate this in brief, let us define the {\it fractional continuum limit} by means of a self-similar function $\Lambda_{a}(h)$ of Weierstrass-Mandelbrot type

\begin{equation}
 \label{selfsim}
\Lambda_{a}(h)=\sum_{s=-\infty}^{\infty} a^{-\delta s}f(a^sh)
\end{equation}
which is self-similar with respect to $h$ fulfilling $\Lambda(ah)=a^{\delta}\Lambda(h)$ and converges for sufficiently good functions $f$. We define the fractional continuum limit by
\begin{equation}
 \label{fraccont}
\lim_{a\rightarrow 1}|\ln(a)|\Lambda_{a}(h) = h^{\delta}\int_0^{\infty}\frac{f(\tau)}{\tau^{\delta+1}}{\rm d}\tau
\end{equation}
which remains finite for $a\rightarrow 1$.
We notice then the following observation: The physics described by the FL defined by the
fractional continuum limit (\ref{fraccont}) of the self-similar Laplacian is up to a prefactor $\sim h^{\delta}$ independent on $h$. The constant $h$ has therefore only the meaning of a dimensional normalization factor (having physical dimension of length), but does not have the meaning of a characteristic length-scale which is nonexistent for self-similar functions fulfilling $\Lambda(ah)=a^{\delta}\Lambda(h)$.
We can perform the fractional continuum limit for any self-similar function of the form (\ref{selfsim}) and the range of convergence for $\delta$ is {\it the same} for both, the
discrete case (\ref{selfsim}) and the fractional continuum limiting case (\ref{fraccont}).

Further illuminating is to consider the fractional continuum limit of a self-similar density function $\rho_{a,\delta}$
generated by Dirac's $\delta$-functions

\begin{equation}
 \label{density}
\rho_{a,h}(\tau) = \lim_{a\rightarrow 1+}\left\{\sum_{s=-\infty}^{\infty} a^{-\delta s} \delta(\tau-ha^{s})\right\} = \frac{h^{\delta}}{\zeta \tau^{\delta+1}}
\end{equation}
where in this limiting case $0<\zeta=|\ln(a)|<<1$ is assumed to be a positive ``small'' but finite quantity ($a\approx 1+\zeta$). The fractional continuum limit of
(\ref{selfsim}) can then be represented by
\begin{equation}
\label{selfsimden}
\Lambda_a(h)= \int_0^{\infty} \rho_{a,h}(\tau)f(\tau){\rm d}\tau
\end{equation}
assuming in the fractional continuum limit the representation (\ref{fraccont}).
Performing the {\it fractional continuum limit for the self-similar Laplacian (\ref{laplacian1D})} defined by

\begin{equation}
\label{limitlap}
 \lim_{a\rightarrow 1}\Delta_{(\delta,a,h)}=\Delta_{(\delta,\zeta)}
\end{equation}
we obtain (up to a positive prefactor) the {\it standard representation of the FL} in one dimension \cite{michela,michelb,michelc}
\begin{equation}
 \label{contilim}
\ds \Delta_{(\delta,\zeta)}u(x) = \frac{h^{\delta}}{2\zeta}\int_{-\infty}^{\infty}\frac{u(x+\tau)+u(x-\tau)-2u(x)}{|\tau|^{\delta+1}}{\rm d}\tau ,\hspace{2cm} 0<\delta<2
\end{equation}
where the additional prefactor $\frac{1}{2}$ comes into play by choosing the integration limits $\pm\infty$ and by taking into account that the integrand (\ref{contilim}) is an even function with respect to $\tau$.
The standard representation of the FL (\ref{contilim}) exists in exactly the same interval $0<\delta<2$ as its discrete self-similar counterpart (\ref{laplacian1D}).
It is now relatively straight-forward to generalize (\ref{contilim}) to the multidimensional case of $n$ dimensions of the physical space

\begin{equation}
 \label{sing1n}
\Delta_{(\alpha,\zeta)} u({\bf x}) = \frac{h^{\alpha}}{2\zeta}\int \frac{u({\bf x}+{\bf r})+
u({\bf x}-{\bf r})-2u({\bf x})}{r^{n+\alpha}}{\rm d}^n{\bf r}       ,\hspace{2cm} 0<\alpha<2,\hspace{0.5cm} n\in \N
\end{equation}
where ${\bf x} \in \R^n$ and the integation is performed over the entire $\R^n$. The generalization
(\ref{sing1n}) to the $\R^n$ includes (up to the prefactors) the standard representation of the FL (\ref{sing1}).
The generalization (\ref{sing1n}) of (\ref{contilim}) maintains the spatial isotropy and the dimensional prefactor $h^{\alpha}$ in (\ref{sing1n}) maintains the physical dimension of the self-similar Laplacian\footnote{whereas the physical dimension of the FL is always $1/(length)^{\alpha}$} (dimension of the energetically conjugate quantity of the field variable) independent of dimension $n$.
The dispersion relation (\ref{dispersion}) takes in the fractional continuum limit the form of a power law \cite{michela}

\begin{equation}
 \label{scalingdis}
\omega_{\delta}^2(k) =A_{\delta}k^{\delta}
\end{equation}
which is in the same time the dispersion relation obtained from the continuum representation (\ref{contilim})
where the constant $A_{\delta}$ is obtained as \cite{michela}

\begin{equation}
 \label{Adelta}
A_{\delta} = \frac{h^{\delta}}{\zeta}\frac{\pi}{\Gamma(\delta+1)\sin{\frac{\delta\pi}{2}}} >0
\end{equation}
where the relation to the normalization factor (\ref{normfac}) (dimension $n=1$) is $A_{\alpha}=\frac{h^{\alpha}}{\zeta}\frac{1}{{\cal C}_{n=1,\alpha}}$.
It is important to note that representation (\ref{sing1n}) exists only for exponents $0<\alpha<2$ for sufficiently regular fields $u({\bf x})$. We emphasize that this restriction is a property of the spatial {\it standard representation} only and not of the operator of FL itself which should exist with all good properties of a true Laplacian for any positive $\alpha>0$.

\subsection{The generalized self-similar potential - towards a generalization of the FL}
\label{gen1}

In this section we introduce a generalization of the above self-similar spring model. Hamilton's variational principle defines then in a rigorous a manner a self-similar Laplacian which takes in the fractional 
continuum limit for fractional non-integer powers new representations of the FL.
To this end it is convenient to introduce the shift operator by

\begin{equation}
 \label{shiftdef}
D_x(h)u(x)=u(x+h)
\end{equation}
where we consider first of all the 1D case.
For sufficiently smooth (infinitely differentiable) fields $u(x)$ where it is sufficient to consider only scalar fields $u$,
the shift operator has the representation

\begin{equation}
 \label{shift}
D_x(h)=\exp{(h\frac{d}{dx})}= \sum_{m=0}^{\infty}\frac{h^m}{m!}\frac{d^m}{dx^m}
\end{equation}
which generates the Taylor series\footnote{We notice that since the shift operator is an operator function of its generator $\frac{d}{dx}$, it has the same eigenfunctions $e^{ikx}$ where  $D_x(h)e^{ikx}= e^{ikh} e^{ikx}$.}.
We notice in brief some essential properties:
$D(h=0)=1$ is the unity operator of zero-shift and $D^{\mu}(h)=D(\mu h)$ for any $\mu\in \R$ and the shift operator is unitary, i.e. $D(-h)=D^{-1}(h)=D_x^{\dagger}(h)$ where with $(..)^{\dagger}$ we denote the adjoint operator of $(..)$.
The shift operation fulfills all properties of a continuous commutative 
Abelian group. 

Let us now introduce a generalized form of discrete self-similar elastic potential (\ref{Welast}). The generalization is such that it involves not only two-point but self-similar harmonic $m+1$-particle interactions which come into play by quadratic forms of finite differences of orders $m=1,2,..\in \N$, namely

\begin{equation}
 \label{Welastgenm}
\ds {\cal W}_m(x,h)=\frac{f_m}{2}\sum_{s=-\infty}^{\infty} a^{-\delta s}\left\{[D_x(ha^s)-1]^mu(x)\right\}^2 ,\hspace{2cm} 0<\delta<2m
\end{equation}
Note that this series is a function of type Weierstrass-Mandelbrot where self-similarity is reflected by \\ ${\cal W}_m(x,ah)=a^{\delta}{\cal W}_m(x,h)$.
This self-similar series converges for sufficiently regular fields $u(x)$ in the interval $0<\delta<2m$. For $m=1$ (\ref{Welastgenm}) coincides with
the potential (\ref{Welast}), i.e. with the self-similar chain model previously introduced \cite{michel}.
Let us again put $f_m=1$ subsequently.
The total elastic energy functional is then given by
\begin{equation}
\label{elastenergytotgnm}
 V_m(h)= \int_{-\infty}^{\infty} {\cal W}(x,h)\,{\rm d}x
\end{equation}

The generalized self-similar potential (\ref{Welastgenm}) fulfills also the self-similarity condition (\ref{self-sim}) and a
generalized self-similar Laplacian is obtained by Hamilton's principle
\begin{equation}
\label{hamilton2m}
 \Delta_{(m,\delta,h)}u(x) =:-\frac{\delta V_m}{\delta u(x)}
\end{equation}

Then this self-similar Laplacian is given by the representation

\begin{equation}
\label{givenres}
\Delta_{(m,\delta,h)}u(x) = \sum_{s=-\infty}^{\infty} a^{-\delta s}
\Delta_{2m}(a^sh) u(x) ,\hspace{2cm} 0<\delta<2m
\end{equation}
where the finite difference operator of even order $2m$ comes into play
\begin{equation}
 \label{lapla1}
\Delta_{2m}(h) = (-1)^{m+1}(D_{x}(\frac{h}{2})-D_{x}(-\frac{h}{2}))^{2m}=-\left(2-D_x(h)-D_x(-h)\right)^m  ,\hspace{2cm} m\in \N
\end{equation}
which is symmetric $\Delta_{2m}(h)=\Delta_{2m}(-h)$ and hence self-adjoint. We restrict ourselves in (\ref{Welastgenm}) on finite differences of integer order $m\in \N$. However, we emphasize that
the entire analysis to follow remains true and can be performed in full explicit form also for finite differences of {\it fractional} orders, where the orders $m$ must be {\it strictly positive $m>0$}. The positiveness is the only real
condition which is necessary to maintain translational invariance  (eigenvalue zero at $k=0$) for the FL to be deduced.
The derivation of (\ref{givenres}) involving operator (\ref{lapla1}) is given in details a recent paper in the context of non-local constitutive models \cite{michel-collet}.
Some important but elementary features of the difference operator which are used to obtain (\ref{lapla1}) are outlined there.

We further notice that for the case $m=1$ the self-similar Laplacian (\ref{laplacian1D}) is re-obtained.
(\ref{givenres}) converges in the same interval $0<\delta<2m$ as the elastic energy density (\ref{Welastgenm}). Taking into account that $e^{ikx}$ are eigenfunctions of the shift operator and hence of (\ref{givenres}) we obtain
the dispersion relation from $\Delta_{(m,\delta,h)}e^{ikx}= - \omega_{m,\delta}^2(kh)e^{ikx}$ in the form of Weierstrass-Mandelbrot type functions
\begin{equation}
\label{disprelamop}
\omega_{m,\delta}^2(kh)= 4^m\sum_{s=-\infty}^{\infty} a^{-\delta s}\sin^{2m}(\frac{kha^s}{2}) ,\hspace{2cm} 0<\delta<2m
\end{equation}
involving powers of orders $2m$ of the sine.
The self-similar Laplacian (\ref{givenres}) as well as the elastic potential
(\ref{Welastgenm}) and the dispersion relation (\ref{disprelamop}) are self-similar functions in the sense of a condition as (\ref{self-sim}). For $m=1$ this model coincides with the self-similar spring model of the previous section leading to
Laplacian (\ref{laplacian1D}). The dispersion relation (\ref{disprelamop}) converges as well in the entire range $0<\delta<2m$ and is a function of the type Weierstrass Mandelbrot which is nowhere differentiable and with fractal features within $0<\delta<1$
(see the figures representing the case $m=1$). The further analysis of the fractal characteristics of Weierstrass-Mandelbrot type functions, however, is not subject of this paper. For further aspects of those functions we refer e.g. to \cite{hardy}.

The fractional continuum limit $a\rightarrow 1+\zeta$ ($0<\zeta<<1$) as introduced in section \ref{fraclim} of
Laplacian (\ref{givenres}) then yields

\begin{equation}
\label{nobu}
\ds \Delta_{m,\delta,\zeta} u(x) = \frac{h^{\delta}}{2\zeta}\int_{-\infty}^{\infty}\frac{\Delta_{2m}(\tau)u(x)}{|\tau|^{\delta+1}}{\rm d}\tau ,\hspace{2cm} 0<\delta<2m
\end{equation}
coinciding with (\ref{contilim}) for $m=1$. For the analysis to follow we will also need its $n$-dimensional counterpart
which has the form
\begin{equation}
 \label{selfsimlap}
\ds \Delta_{m,\delta,\zeta} u({\bf x}) = \frac{h^{\delta}}{2\zeta}\int\frac{\Delta_{2m}({\bf r})u(x)}{r^{\delta+n}}{\rm d}^n{\bf r} ,\hspace{2cm} 0<\delta<2m
\end{equation}
which is well defined in the entire interval $0<\delta<2m$ where the symmetric difference operator $\Delta_{2m}({\bf r})$ is defined below (eq. (\ref{choicem})).
Note that (\ref{nobu}) and (\ref{selfsimlap}) are non-singular integrals which exist within the entire interval $0<\delta<2m$.
The hypersingularity at $\tau=0$ is removed by the tempering property of the difference operator (\ref{lapla1}) behaving as $\sim\tau^{2m}$ for $\tau\rightarrow 0$. 
We will show in the subsequent analysis that (\ref{nobu}) and (\ref{selfsimlap}) represent up to normalization constant the respective 1D and nD FLs for {\it fractional exponents} $0<\alpha<2m$.

However, for integer exponents $\frac{\alpha}{2} \in \N$ the {\it non-regularized expressions} (\ref{nobu}) and (\ref{selfsimlap}) represent integer powers of the Laplacian $\Delta$ only for a restricted subspace of tempered functions
and looses otherwise its analyticity by the appearance of not well defined integrals.
This analytic anomaly at integers merits a deeper analysis which is beyond the scope of this paper. 

Subsequently we deduce a regularized representation of the FL which avoids this anomaly representing for the full range $\frac{\alpha}{2} \geq 0$ the powers 
of the standard Laplacian $\Delta$ including the integer powers $\frac{\alpha}{2} \in \N_0$.

\section{New representations for the FL}

Let us consider an integral of the form
\begin{equation}
\label{rel}
I_{n,\alpha}(k) = \int \frac{f({\bf k}\cdot{\bf r})}{r^{\alpha+n}}{\rm d}^n{\bf r} = I_{n,\alpha}(k=1) k^{\alpha}
\end{equation}
which is performed over the entire $\R^n$ and we notice that it scales as $\sim k^{\alpha}$ for sufficiently good functions $f$
for which this integral exists where the non-zero $I_{n,\alpha}(k=1)$ appears as a normalization constant.
Let us hence assume that $f(\xi)=f(-\xi)$ is an even function.
The integral (\ref{rel}) is regular for functions $f$ of which the hypersingularity $\sim r^{-\alpha-1}$ at $r=0$ is removed.
Such functions require the following asymptotic properties:
\begin{equation}
 \label{asymptotics}
\begin{array}{l}
|f(\xi)| \leq f_0 \xi^{b} ,\hspace{2cm} \xi \rightarrow 0 \nonumber \\
\nonumber \\
|f(\xi)| \leq f_{\infty} \xi^{a} ,\hspace{2cm}  \xi \rightarrow \infty
\end{array}
\end{equation}
where the integral (\ref{rel}) is regular (non-singular at $r=0$) only if $a < \alpha < b$ and $a<b$. The first condition (\ref{asymptotics})$_1$ has in the integral tempering effect at $r=0$ and renders (\ref{rel}) locally integrable at $r=0$.
The second condition (\ref{asymptotics})$_2$ guarantees integrability of (\ref{rel}) at infinity. The conditions (\ref{asymptotics}) define the {\it admissible function space in which (\ref{rel}) is defined}.
In (\ref{asymptotics}) $f_0, f_{\infty} >0$ denote characteristic constants of the function $f$ and where $a<b$ is a necessary condition for $f$ to be admissible.
Now, generate a function
\begin{equation}
 \label{m2}
f_2({\bf h})=\Delta_{2}({\bf h})u(x)
\end{equation}
where $\Delta_{2}({\bf h})$ indicates the symmetric second order finite difference operator of (\ref{lapla1})
for $m=1$, then we notice that $f_2({\bf h})$
fulfills with respect to its dependence on $h$ the properties (\ref{asymptotics}) with $a=0$ and $b=2$. 
Since the standard representation (\ref{sing1}) of the FL is constructed by a second difference operator of the form (\ref{m2}), it indeed is regular within the interval $0<\alpha<2$.

As a generalization we can employ a function $f_{2m} = \Delta_{2m}({\bf r})u({\bf x})$ generated by the symmetric finite
difference operator of order $2m$ and define the integral

\begin{equation}
 \label{integral2morder}
{\cal I}_{m,n,\alpha} = \int \frac{\Delta_{2m}({\bf r})u({\bf x})}{r^{n+\alpha}}{\rm d}^n{\bf r}
\end{equation}
which is regular for $a=0<\alpha=b=2m$. Further it follows from (\ref{rel}) that this integral has for
$u({\bf x})=e^{i{\bf k}\cdot{\bf x}}$ the scaling behavior of (\ref{rel})

\begin{equation}
\label{scabe}
{\cal I}_{m,n,\alpha}(k)= {\cal I}_{m,n,\alpha}(k=1) k^{\alpha} ,\hspace{2cm} 0<\alpha<2m
\end{equation}

The {\it Fractional Laplacian} (FL) $-(-\Delta)^{\frac{\alpha}{2}}$ is defined as the operator which has in Fourier space the (negative definite) representation $-k^{\alpha}$. We will demonstrate that
we can use integrals of the form (\ref{integral2morder}) to derive spatial $\R^n$-representations for the FL which represent powers of the standard Laplacian $\Delta$.
{\it We emphasize that the FL is a uniquely defined operator, so that all spatial representations which generate the FL need to be equivalent and coincide}.

We propose hence the following representation for the FL which will be later justified by its exact definition by Hamilton's variational principle

\begin{equation}
 \label{FL}
-_{m}(-\Delta)^{\frac{\alpha}{2}}u({\bf x}) = {\cal C}_{m,n,\alpha}\int \frac{\Delta_{2m}({\bf r})u({\bf x})}{r^{\alpha +n}}{\rm d}^n{\bf r} ,
 \hspace{2cm} 0<\alpha< 2m ,
\end{equation}
which exists for $0<\alpha< 2m$ for sufficiently regular functions\footnote{i.e. infinitely differentiable and vanishing at infinity.} $u({\bf x})$. In (\ref{FL}) we have employed the symmetric (self-adjoint) finite difference operator
of order $2m$ ($m>0)$, namely
\begin{equation}
 \label{choicem}
\Delta_{2m}({\bf r})= -\left\{2-D_{\bf x}({\bf r})-D_{\bf x}(-{\bf r})\right\}^m=(-1)^{m+1}\left\{D(\frac{\bf r}{2})-D(-\frac{\bf r}{2})\right\}^{2m} ,\hspace{1cm} m \in \N (m>0)
\end{equation}
where $D_{\bf x}$ indicates the unitary shift operator which shifts the vector argument ${\bf x}\rightarrow {\bf x}+{\bf r} $, i.e. $D_{\bf x}({\bf r}) u({\bf x}) = u({\bf x}+{\bf r}) $. The symmetric finite difference operator
(\ref{choicem}) renders (\ref{FL}) locally integrable at $r=0$ due to its tempering behavior $\sim r^{2m}$ for $r$ ``small". The tempering effect is sufficient if $\alpha<2m$ and is present for sufficiently smooth fields. Further we deal with ``good fields" which are vanishing at infinity and as a consequence  $|\Delta_{2m}({\bf r})u(x)| \sim \frac{(2m)!}{m!m!}u({\bf x}) =const$ converges versus a non-zero constant (i.e. independent on $r$) for $\alpha >0$. From these considerations it follows that (\ref{FL}) is a non-singular
integral existing for $0<\alpha<2m$. The normalization constant ${\cal C}_{m,n,\alpha}$ is introduced that (\ref{FL}) has Fourier transform $-k^{\alpha}$.
We show in the appendix \ref{normalization} that this normalization factor is well defined in the entire range $0<\alpha<2m$. However, we will see that it is sufficient to evaluate it explicitly for fractional powers
$\frac{\alpha}{2}\notin \N_0$ only.
\\ \\
{\it We prove subsequently the following features: Expression (\ref{FL}) represents powers of the conventional Laplacian
$\Delta$ only for fractional (non-integer) exponents $\frac{\alpha}{2}\notin \N$. However, in the regularized representations of the FL to be deduced, integer powers
$\frac{\alpha}{2}\in \N$ are also included, including the (due to translational invariance) physically inadmissible value $\alpha=0$. The regularized FL gives a spatial representation in the distributional sense for the powers of the standard Laplacian $-(-\Delta)^{\frac{\alpha}{2}}$ (Equation (\ref{universalFL}) below) which holds for any $\alpha \geq 0$. Especially this regularized representation for the FL is independent of $m$ and coincides with (\ref{FL}) 
in the interval $0< \frac{\alpha}{2}\notin \N< m$.
This regularized representation for the FL takes for integer exponents $\frac{\alpha}{2} \in \N_0$ a distributional representation for the integer powers of the conventional Laplacian $\Delta$ including for $\alpha=0$ the negative unity 
operator.}
\\ \\
We will show now that the FL (\ref{FL}) can be reduced to a {\it unique simple universal representation} which is independent of order $m$. This reduction requires the introduction of a regularization to allow certain manipulations.
This regularization leads to a simple representation of the FL holding for
any $\alpha >0$ representing powers of the standard Laplacian $\Delta$.
To this end we consider the more explicit evaluation of the FL (\ref{FL})

\begin{equation}
 \label{FLexp}
-_m(-\Delta)^{\frac{\alpha}{2}}u({\bf x}) = - {\cal C}_{m,n,\alpha}\int \frac{1}{r^{\alpha +n}}\left\{\frac{(2m)!}{m!m!}u({\bf x})+\sum_{p=1}^m\frac{(2m)!}{(m+p)!(m-p)!}(-1)^p\left[u({\bf x}+p{\bf r})+u({\bf x}-p{\bf r})\right]\right\}{\rm d}^n{\bf r}
\end{equation}
which exists for exponents $0<\alpha< 2m$. The prefactor ${\cal C}_{m,n,\alpha}$ is a normalization constant which
guarantees that the eigenvalues of the FL are given by $-k^{\alpha}$ and is deduced in the appendix
\ref{normalization} and is given by

\begin{equation}
 \label{Amnalphexpli}
 \begin{array}{l}
\ds A_{m,n,\alpha} = {\cal C}_{m,n,\alpha}^{-1}=  U_{n,\alpha}V_{m,\alpha} >0 \nonumber \\ \nonumber \\
\ds \hspace{1cm} = \frac{(-1)^{m+1}\pi^{\frac{n+1}{2}}\Gamma(\frac{\alpha+1}{2})}{2^{\alpha}\Gamma(\frac{\alpha+n}{2})
\Gamma(\alpha+1)\sin{\frac{\alpha\pi}{2}}}
\left(D_{\lambda}(1)-D_{\lambda}(-1)\right)^{2m}|
\lambda|^{\alpha}|_{\lambda=0} ,\,\,0 <\alpha< 2m , \,\,\, \frac{\alpha}{2}\notin \N
\end{array}
\end{equation}
where the integral over the unit-sphere
\begin{equation}
 \label{Uint2}
U_{n,\alpha}=\int_{|\bf n|=1}|n_1|^{\alpha}{\rm d}\Omega({\bf n})=  \frac{2\pi^{\frac{n-1}{2}}\Gamma(\frac{\alpha+1}{2})}{\Gamma(\frac{\alpha+n}{2})} > 0 \hspace{1cm} \alpha >-1
\end{equation}
comes into play and 
where $n_1$ denotes any Cartesian component of the unit-vector ${\bf n}$ and with the integral
\begin{equation}
 \label{Vinteg2}
 \begin{array}{l}
\ds V_{m,\alpha} = 2^{2m-\alpha} \int_0^{\infty}\frac{\sin^{2m}(\xi)}{\xi^{\alpha+1}}{\rm d}\xi >0 \nonumber \\ \nonumber \\
\hspace{1cm} =
\ds (-1)^{m+1}\frac{\pi}{2^{\alpha+1}\alpha!\sin{\frac{\pi}{2}\alpha}}\left(D_{\lambda}(1)-D_{\lambda}(-1)\right)^{2m}|
\lambda|^{\alpha}|_{\lambda=0}
,\hspace{1cm} 0<\frac{\alpha}{2} \notin \N <m
\end{array}
\end{equation}

Note that relation (\ref{Vinteg2})$_2$ holds only for non-integers $0< \frac{\alpha}{2} \notin \N < m$.

{\it It is crucially important to note that (\ref{FLexp}) does not allow commutation of the binomial sum and ${\bf r}$-integration,
since the ${\bf r}$-integrals become singular without the {\it tempering effect at $r=0$} of the $\Delta_{2m}$-difference operator\footnote{$\Delta_{2m}({\bf r}) \sim r^{2m}$ as $r\rightarrow 0$)}.
In order to allow this commutation integral (\ref{FLexp}) has to be regularized}. We observe that the integrand is an even function in ${\bf r}$.
By introducing polar coordinates ${\bf r}={\bf n}r$ (${\bf n}$ denotes the unit-vector in ${\bf r}$-direction)
(\ref{FLexp}) can be further written as

\begin{equation}
 \label{expre}
-_m(-\Delta)^{\frac{\alpha}{2}}u({\bf x}) = - {\cal C}_{m,n,\alpha}\int_{|{\bf n}|=1}{\rm d}\Omega({\bf n})
\left(\frac{(2m)!}{m!m!}\int_0^{\infty}{\rm d}r\frac{u({\bf x})}{r^{\alpha+1}}+ 2\sum_{p=1}^m \frac{(2m)!}{(m+p)!(m-p)!}(-1)^p\int_0^{\infty}{\rm d}r\frac{u({\bf x}+pr{\bf n})}{r^{\alpha +1}}\right)
\end{equation}
We emphazize that by commuting integration and summation in (\ref{expre}), the termpering effect of the $\Delta_{2m}$-difference operator does not take place, and hence (\ref{expre}) is a sum of
{\it singular integrals at $r=0$} requiring regularization in order to be defined. The regularization needs to
eliminate the divergence of any integral term of the form
\begin{equation}
 \label{divergence}
|\int \frac{{\rm d}^n{\bf r}}{r^{n+\alpha}}u({\bf x}+p{\bf r})| = |p|^{\alpha} |\int \frac{{\rm d}^n{\bf r}}{r^{n+\alpha}}u({\bf x}+{\bf r})|\rightarrow \infty ,\hspace{1cm} \alpha >0 
\end{equation}
which diverges due to the hypersingularity of order $ \sim r^{-\alpha-1}$ of the integrand at $r=0$. To define (\ref{expre}) we need a regularization
with the following properties:\newline\newline
\noindent(i) {\it The regularization of (\ref{expre}) after the commutation coincides for good fields $u$ with the unregularized well defined representation (\ref{FLexp})} \newline
\noindent(ii) {\it The regularization renders (\ref{divergence}) finite in order to allow commutation of integration
and summation in (\ref{expre})}.
\\ \\
We employ the following regularization rule that fulfills criteria (i),  and (ii) which was already employed earlier  \cite{gelfand,michela,michelb,michelc,michel-fcaa}

\begin{equation}
\label{PV}
-|\xi|^{-\alpha-1} \sin{\frac{\pi\alpha}{2}} = \lim_{\epsilon\rightarrow 0+}\Re(\epsilon-i\xi)^{-\alpha-1} , \hspace{2cm}
\end{equation}
where $\Re(..)$ denotes the real-part of $(..)$. This relation is an exact equality for non-zero $|\xi| > 0$. For $\xi=0$ the right hand side guarantees local integrability 
at $\xi=0$ by removing the hypersingularity of the left hand
side which occurs when $\alpha>0$. We can say that regularization rule (\ref{PV}) is an equation in the distributional sense, i.e. it holds ``under the integral''.
The regularizing effect of (\ref{PV}) can be demonstrated for instance by considering the integral ($\xi_0>0$) and $\alpha > 0$
\begin{equation}
 \label{intrgdemo}
  I_{reg}(\xi_0,\alpha)=\lim_{\epsilon\rightarrow 0+}\Re\int_0^{\xi_0}(\epsilon-i\xi)^{-\alpha-1}{\rm d}\xi = \frac{\sin{\frac{\pi\alpha}{2}}}{\alpha}\xi_0^{-\alpha}
\end{equation}

Integrating the left hand side of (\ref{PV}) yields a singular integral with a diverging term $-\frac{\sin{\frac{\pi\alpha}{2}}}{\alpha}\xi^{-\alpha} $ ($\xi\rightarrow 0$) 
at the 
lower integration limit $\xi=0$. The crucial feature of regularization rule (\ref{PV}) is that the divergence of the lower limit is removed leading to the finite value 
(\ref{intrgdemo}).
Furthermore (\ref{PV}) gives for integer $\frac{\alpha}{2} \in \N_0$ to the left hand side 
a distributional meaning by means even derivatives of Dirac's $\delta$-function. That is, by employing regularization rule (\ref{PV}) integrals containing
$-|\xi|^{-\alpha-1} \sin{\frac{\pi\alpha}{2}}$ are well defined. We will show subsequently that the FL (\ref{expre}) contains such term and hence 
the FL (\ref{expre}) is well defined also for integers $\frac{\alpha}{2} \in \N_0$ if regularization rule (\ref{PV}) is applied and by accounting its distributional
form for these cases
\begin{equation}
 \label{notexpreg}
-\frac{\sin{\frac{\pi\alpha}{2}}}{|\xi|^{\alpha+1}} = \lim_{\epsilon\rightarrow 0+}
  \Re\{\frac{i^{\alpha+1}}{(\xi+i\epsilon)^{\alpha+1}}\} =(-1)^{p}\frac{\pi}{(2p)!}\frac{d^{2p}}{d\xi^{2p}}\delta(\xi) ,\hspace{0.2cm}  \frac{\alpha}{2}=p \in \N_0
\end{equation}
where $\delta(\xi)$ denotes Dirac's $\delta$-function. It follows that in the limiting case of integer exponents
(\ref{notexpreg}) takes completely local distributional representations of even-oder derivatives.
Further we notice that in the limiting case $\alpha\rightarrow 0+$ the regularized integral (\ref{intrgdemo}) tends to a finite value $I_{reg} \rightarrow \frac{\pi}{2}$.

We hence can write for (\ref{expre}) which is defined only by employing regularization rule (\ref{PV}) 
\begin{equation}
 \label{exprereg}
\begin{array}{l}
\ds -_m(-\Delta)^{\frac{\alpha}{2}}u({\bf x}) = \nonumber \\ \nonumber \\
 - {\cal C}_{m,n,\alpha}\int_{|{\bf n}|=1}{\rm d}\Omega({\bf n})
\ds \left(\frac{(2m)!}{m!m!}u({\bf x})\int_0^{\infty}\frac{{\rm d}\xi}{\xi^{\alpha+1}} + 2\int_0^{\infty}
\frac{u({\bf x}+\xi{\bf n})}{\xi^{\alpha+1}}{\rm d}\xi \sum_{p=1}^m \frac{(2m)!}{(m+p)!(m-p)!}(-1)^p  p^{\alpha}\right)
\end{array}
\end{equation}

In this way we can
for the terms $p\neq 0$ separately re-scale the integration variables to $\xi=pr$ and obtain for all terms of $p\neq 0$ the same integral with integrand $\frac{u({\bf x}+\xi{\bf n})}{\xi^{\alpha+1}}$. The term for $p=0$ contributes only as integral of $\frac{1}{\xi^{\alpha+1}}$.
This term $p=0$ is vanishing in the regularization (leading to an integral of the form (\ref{intrgdemo}) with $\xi_0\rightarrow \infty$).
So the constant term $\sim u({\bf x})$ which corresponds to $p=0$ does not contribute. Hence we can rewrite the sum which can be extracted from relation (\ref{exprereg}) as\footnote{with the shift operator $D_{\lambda}(a)f(\lambda)|_{\lambda=0}=f(\lambda+a)|_{\lambda=0}=f(a)$.}
\begin{equation}
\label{zeroterm}
\begin{array}{l}
\ds 2\sum_{p=1}^m \frac{(2m)!}{(m+p)!(m-p)!}(-1)^p p^{\alpha} =\frac{(2m)!}{m!m!}0^{\alpha}+ 2\sum_{p=1}^m \frac{(2m)!}{(m+p)!(m-p)!}(-1)^p p^{\alpha} \nonumber \\ \nonumber \\
\ds =\frac{(-1)^m}{2^{\alpha}}\left[D_{\lambda}(1)-D(-1)\right]^{2m}|\lambda|^{\alpha}|_{\lambda=0} ,\hspace{2cm} \alpha > 0
\end{array}
\end{equation}
where the vanishing term of $p=0$ has been added for convenience.
Note that this relation is in the range $0<\frac{\alpha}{2} < m$ {\it non-vanishing only for $\frac{\alpha}{2} \notin \N$}.

We obtain then for the FL the following (without regularisation rule (\ref{PV}) singular) representation 

\begin{equation}
 \label{reguFL}
\begin{array}{l}
\ds -_m(-\Delta)^{\frac{\alpha}{2}}u({\bf x}) = \nonumber \\ \nonumber \\
\ds (-1)^{m+1}{\cal C}_{m,n,\alpha}2^{-\alpha}\left[D_{\lambda}(1)-D(-1)\right]^{2m}|\lambda|^{\alpha}|_{\lambda=0}
\int\frac{{\rm d}^n{\bf r}}{r^{n+\alpha}}u({\bf x}+{\bf r})  \nonumber \\ \nonumber \\\ds  =
\frac{2\alpha!\sin{\frac{\alpha\pi}{2}}}{\pi U_{n,\alpha}} \int\frac{{\rm d}^n{\bf r}}{r^{n+\alpha}}u({\bf x}+{\bf r}) ,\hspace{2cm} \alpha>0 \nonumber \\ \nonumber \\ \ds 
-(-\Delta)^{\frac{\alpha}{2}} u({\bf x}) = C_{n,\alpha}\int \frac{u({\bf r})}{|{\bf x}-{\bf r}|^{n+\alpha}}{\rm d}^n{\bf r} ,\hspace{2cm} \alpha >0
\end{array}
\end{equation}
which is independent of $m$ and well defined in the distributional sense when employing regularization rule (\ref{PV}).
The normalization constant which is indepedent of $m$ is then obtained by
\begin{equation}
 \label{yeilds}
C_{n,\alpha}=\frac{2\alpha!\sin{\frac{\alpha\pi}{2}}}{\pi U_{n,\alpha}} = \frac{\Gamma(\frac{\alpha+n}{2})\alpha!\sin{\frac{\alpha\pi}{2}}}{\pi^{\frac{n+1}{2}}\Gamma(\frac{\alpha+1}{2})}
\end{equation}
coinciding for $0<\alpha<2$ with the normalization factor of the standard representation (\ref{normfac}).  Then we obtain the {\it regularized representation} for the FL 
\begin{equation}
 \label{universalFL}
 \begin{array}{l}
 \ds -(-\Delta)^{\frac{\alpha}{2}} u({\bf x}) = -\frac{2 \alpha !}{\pi U_{n,\alpha}}\lim_{\epsilon\rightarrow 0+}\Re\left\{i^{\alpha+1}\int \frac{u({\bf x}+{\bf r})}{(r+i\epsilon)^{\alpha+1}}\frac{{\rm d}^n}{r^{n-1}}\right\} \nonumber \\  \nonumber \\
\ds -(-\Delta)^{\frac{\alpha}{2}} u({\bf x}) = -\frac{\Gamma(\frac{\alpha+n}{2})\alpha!}{\pi^{\frac{n+1}{2}}\Gamma(\frac{\alpha+1}{2})}
\lim_{\epsilon\rightarrow 0+}\Re\left\{i^{\alpha+1}\int\frac{u({\bf r})}{(|{\bf x}-{\bf r}|+i\epsilon)^{\alpha+1}}\frac{{\rm d}^n{\bf r}}{|{\bf x}-{\bf r}|^{n-1}} \right\} ,\hspace{2cm} \alpha \geq 0
\end{array}
\end{equation}

We emphasize that (\ref{universalFL}) holds for all positive exponents $\alpha \geq 0$ including integers (with zero) that (\ref{universalFL}) represents the powers of the convential Laplacian $\Delta$.
We notice that for fractional powers $\frac{\alpha}{2} \notin \N_0$, the regularized FL (\ref{universalFL}) is a non-local convolutional (scale-free and self-similar) operator which remains non-local whatever the scale of $r$ representation may be, whereas
(\ref{universalFL}) becomes a local operator for integer powers $\frac{\alpha}{2} \in \N_0$. (\ref{universalFL})
can be considered as {\it is the analytical extension} of the FL (\ref{FL}) to full range of exponents $\alpha \geq 0$.

For the 1D case $n=1$ the regularized FL of (\ref{notexpreg}), (\ref{universalFL}) yields

\begin{equation}
 \label{1Drep}
\begin{array}{l}
\ds -(-\frac{d^2}{dx^2})^{\frac{\alpha}{2}}u(x)=  \lim_{\epsilon\rightarrow 0+}
-\frac{\alpha!}{\pi} \Re\left\{i^{\alpha+1}\int_{-\infty}^{\infty}\frac{u(\tau)}{(x-\tau+i\epsilon)^{\alpha+1}}{\rm d}\tau\right\} ,\hspace{2cm} \alpha \geq 0
\ds \nonumber \\  \nonumber \\
\ds \hspace{0.5cm} =: \frac{\alpha!\sin{\frac{\pi\alpha}{2}}}{\pi}\int_{-\infty}^{\infty}\frac{u(x+\tau)}{|\tau|^{\alpha+1}}{\rm d}\tau ,\hspace{2cm} \alpha >0 ,\frac{\alpha}{2} \notin \N
\end{array}
\end{equation}
where integers $\frac{\alpha}{2}=p\in \N_0$ yield the required result $-(-\frac{d^2}{dx^2})^p$. Relation (\ref{1Drep})$_2$ is its unregularized singular representation which is
defined {\it only} in the sense of (\ref{1Drep})$_1$, i.e. by employing regularization (\ref{PV}).

Let us now analyze in full details the case of $n$ dimensions for $\frac{\alpha}{2}=p\in \N_0$ to prove that for $n$ dimensions the regularized representation of the FL 
(\ref{universalFL}) indeed yields for $\frac{\alpha}{2}=p\in \N_0$ the integer powers 
$-(-\Delta)^p$ of the conventional Laplacian.
To this end we can write for (\ref{universalFL}) by using the shift operator (\ref{shift}) where only the even powers contribute\footnote{The gradient $\nabla_{\bf x}$ acts only on the ${\bf x}$-dependence of $u({\bf x})$.}

\begin{equation}
 \label{caslimite2p}
-(-\Delta)^{\frac{\alpha}{2}=p}u({\bf x})= -\frac{1}{ U_{n,2p}}\sum_{s=0}^{\infty}\int_{|{\bf n}|=1}\frac{({\bf n}\cdot\nabla_{\bf x})^{2s}u({\bf x})}{(2s)!}{\rm d}\Omega({\bf n})\Re\left\{\frac{2(2p)! \, i^{2p+1}}{\pi}\int_0^{\infty}\frac{r^{2s}}{(r+i\epsilon)^{2p+1}}{\rm d}r\right\}
\end{equation}
where $U_{n,2p}$ is the surface integral
(\ref{Uint2}) for even integers $\alpha=2p \in \N_0$. Further we observe by taking into account relation  (\ref{1Drep}) that the $r$-integral yields
after $2p$ partial integrations

\begin{equation}
\label{observe}
\begin{array}{l}
\ds 2 \Re\left\{\frac{(2p)!i^{2p+1}}{\pi}\int_0^{\infty}\frac{r^{2s}}{(r+i\epsilon)^{2p+1}}{\rm d}r\right\}  \nonumber \\ \nonumber \\  \ds = 2(-1)^p\int_0^{\infty}r^{2s}\frac{d^{2p}}{dr^{2p}}\delta(r){\rm d}r  \nonumber \\ \nonumber \\
\ds = 2 (-1)^p (2p)! \delta_{ps} \int_0^{\infty}\delta(r){\rm d}r = (-1)^p (2p)!\delta_{ps}
\end{array}
\end{equation}
being non-zero only for the order $2s=2p$ ($\delta_{ps}$ denotes the Kronecker-$\delta$). With this relation we can write for
(\ref{caslimite2p})

\begin{equation}
\label{writeFFL}
 -(-\Delta)^{p}\, u({\bf x}) =   \frac{(-1)^{p+1}}{U_{n,2p}}\int_{|{\bf n}|=1}({\bf n}\cdot\nabla_{\bf x})^{2p}{\rm d}\Omega({\bf n}) \, u({\bf x})
\end{equation}

To evaluate this surface integral, we utilize (\ref{Uint2}) to arrive at

\begin{equation}
\label{mono2p}
\begin{array}{l}
 \int_{|{\bf n}|=1}{\rm d}\Omega({\bf n})({\bf a}\cdot{\bf n})^{2p} = U_{n,\alpha=2p} ({\bf a}\cdot{\bf a})^p\nonumber \\
\nonumber \\
=U_{n,\alpha=2p} \sum_{p_1+p_2+..p_n=p} \frac{(2p)!}{(2p_1)!(2p_2)!..(2p_n)!}a_1^{2p_1}a_2^{2p_2}..a_n^{2p_n}
\end{array}
\end{equation}
where ${\bf a}$ denotes a constant vector (independent on the surface unit-vector ${\bf n}$). In (\ref{mono2p}) only even combinations of powers of the
Cartesian components $n_i$ of the unit vector yield non-vanishing contributions.
Replacing in this relation the monomials by partial derivatives $a_1^{2p_1}a_2^{2p_2}..a_n^{2p_n} =\frac{\partial^{2p}}{\partial x_1^{2p_1}\partial x_2^{2p_2}..\partial x_n^{2p_n}}$ gives for (\ref{writeFFL}) finally

\begin{equation}
\label{final2p}
 -(-\Delta)^{p}\, u({\bf x}) = (-1)^{p+1}(\nabla\cdot\nabla)^{2p} ,\hspace{1cm} p\in \N_0
\end{equation}
which indeed is the correct integer powers of the standard Laplacian including for the case $p=0$ where (\ref{final2p}) yields $-u({\bf x})$ since $(-\Delta)^{p=0}=1$ yields the unity operator.
The sign $(-1)^{p+1}$ alternating with order $p$ appears always such that (\ref{final2p}) has Fourier representation $-k^{2p}$ with all good properties of a Laplacian.

\subsection{Hamilton's variational principle: the FL versus self-similar Laplacian}

In this section we demonstrate that harmonic self-similar elastic energy density functionals of Weierstrass-Mandelbrot function type lead via application of the fractional continuum limit defined in section \ref{fraclim} together with Hamilton's variational principle to rescaled versions of the non-regularized FL of the type (\ref{FL}) with limited intervals of existence $0<\alpha<2m$. Application of regularization rule
(\ref{PV}) then allows to arrive at the {\it universal regularized fractional Laplacian FL (\ref{universalFL})}. We show this to be true for any admissible self-similar harmonic potential energy densities of Weierstrass-Mandelbrot type.
In this way we demonstrate the general interlink between self-similar elastic potentials
(describing physically the absence of an intrinsic lengthscale  of the range of interparticle interactions), and the appearance of the regularized FL (\ref{universalFL}) as an universal footprint of self-similar elastic systems.
Our goal is to demonstrate in this section that the physical information on the details of the harmonic interactions of the self-similar potentials appears only as a {\it characteristic scaling factor in the rescaled FL}. In other words: The physics on such self-similar harmonic systems is independent on the details of the self-similar potential.
This characteristic scaling factor does not contain any
absolute length scale being again a consequence of self-similarity (absence of internal length scale of the interparticle interactions).

Let us consider a discrete elastic energy density of self-similar {\it Weierstrass-Mandelbrot function type} (\ref{Welastgenm}) which can be written as

\begin{equation}
 \label{1D}
{\cal W}_a(x,h)=\frac{1}{2}\sum_{s=-\infty}^{\infty}a^{-\alpha s}\left\{(D_x(ha^s)-1)^mu(x)\right\}^2 ,\hspace{2cm} 0<\alpha<2m
\end{equation}
which fulfills the self-similarity condition ${\cal W}(ah)= a^{\alpha}{\cal W}(h)$.
Application of Hamilton's variational principle to (\ref{1D}) yields the self-similar discrete Laplacian (\ref{givenres}) which is also of Weierstrass-Mandelbrot type.
The {\it fractional continuum limit $a\rightarrow 1$} of (\ref{1D})
yields by using (\ref{fraccont}) the 1D integral ($a=1+\zeta$, $0<\zeta \approx ln(a) <<1$)
\begin{equation}
\label{1Dfractui}
\ds {\cal W}_{a=1+\zeta}(x,h) = \frac{h^{\alpha}}{2\zeta}\int_0^{\infty}\frac{{\rm d}\tau}{\tau^{\alpha+1}}
\left\{(D_x(\tau)-1)^mu(x)\right\}^2 ,\hspace{2cm} 0<\alpha<2m 
\end{equation}
The total elastic energy is then given by
\begin{equation}
\label{totalela}
\begin{array}{l}
\ds V_{m,1,\alpha} =  \int_{-\infty}^{\infty}{\cal W}_{a=1+\zeta}(x,h){\rm d}x \nonumber \\ \nonumber \\
\ds V_{m,1,\alpha} = \frac{h^{\alpha}}{4\zeta}\int_{-\infty}^{\infty}\int_{-\infty}^{\infty}\frac{{\rm d}\tau{\rm d}x}{\tau^{\alpha+1}}
\left\{(D_x(\tau)-1)^mu(x)\right\}^2
\end{array}
\end{equation}
where the additional prefactor $\frac{1}{2}$ comes into play by taking both integrals in the limits $(-\infty,\infty)$.
The generalization of the 1D self-similar elastic energy (\ref{totalela}) to the $n$-dimensional space yields the representation

\begin{equation}
 \label{elastener1}
V_{m,n,\alpha}[u] = \frac{h^{\alpha}}{4\zeta}\int \int
\frac{\left\{(D_{\bf x}({\bf r})-1)^mu({\bf x})\right\}^2}{r^{\alpha +n}}{\rm d}^n{\bf r}{\rm d}^n{\bf x} ,
 \hspace{2cm} 0<\alpha< 2m
\end{equation}
Application of Hamilton's variational principle to (\ref{elastener1}) yields the self-similar continuous Laplacian (\ref{selfsimlap}) which is up to its different prefactor coinciding with the {\it non-regularized} FL (\ref{FL}).
The positiveness of this normalization factor (\ref{Amnalphexpli}) appearing in (\ref{FL})
reflects elastic stability (positive definiteness of the self-similar potential).  

Performing the procedure of the last section toegther with regularization rule (\ref{PV}) finally yields
a rescaled version of the regularized FL (\ref{universalFL}).

It is noteworthy that, the normalization constant (\ref{yeilds}) of the regularized  FL is due to the oscillations of $\sin{\frac{\alpha\pi}{2}}$ alternating in sign which indicates that the regularized representation of the FL (\ref{universalFL})
cannot be derived {\it directly} from a variational principle as a functional derivative of a potential such as (\ref{elastener1}).

Now let us analyze a generalization of the elastic energy functional (\ref{elastener1}) in the following form\footnote{Where the symmetric constant matrix $V_{pq}=V_{qp}=V_{|p-q|}$
having only one single eigenvalue zero, due translational invariance with $V(u=const)=0$.}

\begin{equation}
 \label{enfu}
V_{\alpha}[u] = \frac{1}{4}\int\int\frac{{\rm d}^n{\bf y}{\rm d}^n{\bf r}}{r^{n+\alpha}}\sum_{p,q}V_{|p-q|}u({\bf y}+p{\bf r})u({\bf y}+q{\bf r}) \geq 0 ,\hspace{1cm} \hspace{0.5cm} 0<\alpha< \beta
\end{equation}
where we assume that (\ref{enfu}) is well defined for $0<\alpha<\beta$ where $\beta$ depends on the choice of the matrix $V_{|p-q|}$. Further we have
$V_{\alpha}[u=const]= 0$ which indicates translational invariance (zero elastic energy for uniform translations $u=const$\footnote{where $=const$ is not a ``good field" in the sense of (\ref{asymptotics})} and $V_{\alpha}[u] >0$ otherwise).
The self-similar continuous ``Laplacian'' $\Delta_{V}$ of this system is defined by Hamilton's variational principle leading to\footnote{where we write here just the integrals in their singular representations which are to be defined by applying regularization rule (\ref{PV}).}

\begin{equation}
\label{LaplacianV}
\begin{array}{l}
\ds \Delta_{V}u({\bf x})= - \frac{\delta }{\delta u({\bf x})}V_{\alpha}[u] \nonumber \\ \nonumber \\
\ds \Delta_{V}u({\bf x}) = -\frac{1}{4}\int\frac{{\rm d}^n{\bf r}}{r^{n+\alpha}}\sum_{p,q}V_{pq}\left(u({\bf x}+(p-q){\bf r}))+
u({\bf x}-(p-q){\bf r})\right) \nonumber \\ \nonumber  \\
\ds \Delta_{V}u({\bf x}) = -\frac{1}{2}\int\frac{{\rm d}^n{\bf r}}{r^{n+\alpha}}\sum_{p,q}V_{pq}u({\bf x}+(p-q){\bf r})
\end{array}
\end{equation}

To evaluate last step (\ref{LaplacianV})$_3$ we face the same divergence problem as in (\ref{divergence}) when we commute summation and integration. To render this commutation admissible we need to apply regularization rule (\ref{PV}) which regularizes each singular integral term of the sum.
Then (\ref{LaplacianV})$_3$ takes the representation

\begin{equation}
\label{adminreg}
\ds \Delta_Vu({\bf x}) = -\frac{1}{2}\sum_{pq}V_{pq}\int\frac{{\rm d}^n{\bf r}}{r^{n+\alpha}}u({\bf x}+(p-q){\bf r}) 
\end{equation}
which is as relation (\ref{1Drep})$_2$ a sum of singular integrals, and only defined by employing regularization rule (\ref{PV}).
By plugging in (\ref{PV}) into (\ref{adminreg}) we can integrate any term of the sum by rescaling its radial integration variable 
(by introducing for each term $p\neq q$ the new intgration variable 
$|p-q|r=\xi \rightarrow r$) and with the same reasoning which lead us to to above relation (\ref{exprereg}), 
we can evaluate each term of the sum (\ref{adminreg}) by using

\begin{equation}
\label{rescaleit}
\int\frac{u({\bf x}+(p-q){\bf r})}{r^{n+\alpha}}{\rm d}^n{\bf r} = |p-q|^{\alpha}\int\frac{u({\bf x}+{\bf r})}{r^{n+\alpha}}{\rm d}^n{\bf r}
\end{equation}
This relation is for $\alpha >0$ also valid for $p=q$, since the regularized value of the integrals occuring at $p=q$ are vanishing due to (\ref{intrgdemo})
\begin{equation}
\label{regulval}
 \int\frac{{\rm d}^n{\bf r}}{r^{n+\alpha}} = 0
 \end{equation}
So we obtain for the Laplacian (\ref{LaplacianV}) just a {\it rescaled version of the FL (\ref{reguFL})}, namely

\begin{equation}
 \label{laplacianfinal}
\Delta_{V}u({\bf x}) = - {\cal A}_{V}\int\frac{u({\bf x}+{\bf r})}{r^{n+\alpha}}{\rm d}^n{\bf r}
\end{equation}
being defined only with regularization rule (\ref{PV}) and in this way
being a rescaled version
of the regularized FL (\ref{universalFL}). We emphasize that (\ref{laplacianfinal}) is {\it only
defined in regularized sense}, i.e. by plugging in regularization rule (\ref{PV}).
The characteristic {\it positive} scaling factor
\begin{equation}
\label{scalingfactor}
{\cal A}_{V}= \frac{1}{2}\sum_{p,q}V_{|p-q|}|p-q|^{\alpha} >0 , \hspace{1.5cm} \alpha >0 
\end{equation}
can be conceived as a self-similar stiffness matrix containing the ``scale free" information of the matrix $V_{|p-q|}$, but it does not contain any information on an absolute length-scale of the interparticle interactions. 
This is expressed by the self-similar power functions $|p-q|^{\alpha}$:
Any change of scale of $(p-q)\rightarrow \lambda(p-q)$ only rescales the characteristic scaling factor (\ref{scalingfactor}) by $|\lambda|^{\alpha}$.
For the special potential (\ref{elastener1}) the scaling factor (\ref{scalingfactor}) is evaluated in the appendix \ref{normalization}.
From this demonstration follows that the regularized representation (\ref{universalFL}) of the FL is the universal and unique charactersitic operator of all harmonic self-similar systems with elastic potentials of the general form (\ref{enfu}). Note that the scaling factor (\ref{scalingfactor}) is only well defined for the physically admissible exponents $\alpha >0$ which reflects its ``physical origine" (respecting translational invariance) whereas the regularized FL (\ref{universalFL}) is mathematically well defined in the distributional sense for $\alpha \geq 0$
including also the physically forbidden value $\alpha=0$.

\section{Conclusions}

The main goal of the present analysis was to deduce a spatial distributional representation for the FL $-(-\Delta)^{\frac{\alpha}{2}}$ which exists for any $\alpha \geq 0$ (Eq. (\ref{universalFL})). The deduction of this reprepresentation inevitably requires regularization (relation (\ref{PV})).
We analyzed self-similar elastic potentials of the form (\ref{enfu}) which include potentials (\ref{elastener1}) that are obtained by the
fractional continuum limit from discrete self-similar Weierstrass-Mandelbrot type potentials of the form (\ref{1D}). Hamilton's variational principle then defines in a rigorous manner self-similar
Laplacians (\ref{laplacianfinal}) and (\ref{selfsimlap}), respectively which take the representations of rescaled variants of the universal regularized FL (\ref{universalFL}).
Their characteristic scaling factors contain the information on the self-similar elastic potentials (stiffness matrices). Since the characteristic information on the elastic potentials
appears only in the form of different scaling factors of the regularized FL (\ref{universalFL}), this operator has universal character for {\it harmonic systems with self-similar, scale-free interactions}.
All these self-similar harmonic systems are uniquely governed by the {\it regularized FL (\ref{universalFL})}.
The appearence of this universal regularized representation of the FL opens a considerable reduction of difficulty in order to handle physical problems which are described in a fractional calculus context. The extended range of validity to exponents $\alpha \geq 0$ especially opens newly emerging problems to the analysis which are described by exponents outside the L\'evy-interval $(0,2)$ such as recently described for distributions of fluctuations in financial market indices \cite{gopikrishnan}.

The self-similar spring model invoked as point of departure for our approach as well as the generalized discrete self-similar potentials of Weierstrass-Mandelbrot type (\ref{1D}) provide the interlink to the inherent
{\it fractal properties} of these systems, having as a consequence dispersion relations (\ref{disprelamop}) of the type of Weierstrass-Mandelbrot functions with fractal features in the regime of
exponents $0<\delta<1$. In this way we have demonstrated the relation between fractality of 
certain physical properties such as of the Weierstrass-Mandelbrot type dispersion relation (as visualized in the figures) and the ``generalized'' fractional Laplacian approach presented in this paper.
There is a big deal of interdisciplinary scaling invariant problems in various physical contexts such as in critical phenomena, anomalous diffusion and related 
stochastic motions such as in turbulence, where the regularized representation of the FL (\ref{universalFL}) and its links to Weierstrass-Mandelbrot fractal functions, 
may be useful. 
We hope the present paper inspires further works on those and related problems.

\section{Acknowledgements}
Je suis Charlie.

\begin{appendix}
\section{Appendix}
\label{append}

\subsection{Explicit determination of the normalization constant ${\cal C}_{m,n,\alpha}$}
\label{normalization}

In this appendix we determine the normalization constant ${\cal C}_{m,n,\alpha}$ which comes into play
to normalize the representation of the FL (\ref{FL}) such that its Fourier transform is $-k^{\alpha}$.
in explicit form. The normalization constant is determined by the exigence that the FL has Fourier the transform $-k^{\alpha}$.
Accounting for this, it follows from (\ref{scabe}) that

\begin{equation}
 \label{norcon}
{\cal C}_{m,n,\alpha}\cdot {\cal I}_{m,n,\alpha}(k=1)=-1
\end{equation}

For our convenience we put $-{\cal I}_{m,n,\alpha}(k=1)=A_{m,n,\alpha}={\cal C}_{m,n,\alpha}^{-1} $ which takes the form
\begin{equation}
 \label{basishelpintegral}
A_{m,n,\alpha} =(-1)^m\int \frac{(e^{i\frac{{\hat k}\cdot{r}}{2}}-e^{-i\frac{{\hat k}\cdot{r}}{2}})^{2m}}{r^{n+\alpha}}{\rm d}^n{\bf r} = 2^{2m-\alpha}\int \frac{\sin^{2m}({\hat k}\cdot{\bf r})}{r^{n+\alpha}}{\rm d}^n{\bf r} >0 ,
\hspace{1cm} 0< \alpha < 2m
\end{equation}
where  ${\hat k}\cdot{\hat k}=1$.
The constant $A_{m,n,\alpha} >0$ is positive. Its positiveness is crucial for the FL in order to have all good properties of a Laplacian. We notice that the constant (\ref{basishelpintegral}) is defined for any $0<\alpha<2m$, i.e. also for $\frac{\alpha}{2} \in \N$.
By introducing the polar coordinates ${\bf r}={\bf n}$ ($r=|{\bf r}|$ and ${\bf n}\cdot{\bf n}=1$) and choosing the coordinate system
such that ${\hat k}\cdot{\bf r}=rn_1$ where $n_1$ is a director cosine. We notice that
(\ref{basishelpintegral}) can be factorized according to ($A_{m,n,\alpha}={\cal C}_{m,n,\alpha}^{-1}$)

\begin{equation}
 \label{basishelpintegral2}
A_{m,n,\alpha} ={\cal C}_{m,n,\alpha}^{-1} = U_{n,\alpha}V_{m,\alpha} ,\hspace{2cm} 0<\alpha<2m
\end{equation}

with the surface integral over the unit sphere

\begin{equation}
 \label{Uint}
U_{n,\alpha}=\int_{|\bf n|=1}|n_1|^{\alpha}{\rm d}\Omega({\bf n})=  \frac{2\pi^{\frac{n-1}{2}}\Gamma(\frac{\alpha+1}{2})}{\Gamma(\frac{\alpha+n}{2})} > 0
\end{equation}

This integral $U_{n,\alpha}$ exists for $\alpha >-1$ and is explicitly evaluated in the subsequent appendix 
\ref{norm-fac}.
For instance (\ref{Uint}) takes in 1D the value $U_{n=1,\alpha}=2$.
In (\ref{basishelpintegral2}) we further have the integral

\begin{equation}
 \label{Vinteg}
V_{m,\alpha} = 2^{2m-\alpha} \int_0^{\infty}\frac{\sin^{2m}(\xi)}{\xi^{\alpha+1}}{\rm d}\xi >0 ,\hspace{1cm} 0<\alpha<2m
\end{equation}

The existence of (\ref{Vinteg}), especially  $0< V_{m,\alpha}< \infty$ and hence $ A_{m,n,\alpha}>0$ which indicates ``elastic stability'' (positive definiteness) of the system which is due to the fact that (\ref{selfsimlap}) is the outcome of Hamilton's variational pronciple of a positive definite elastic potential in the entire open interval $0<\alpha<2m$. A further point is crucial: 
We see in view of (\ref{Uint}) and (\ref{Vinteg}) that the normalization factor
as well as (\ref{FL}) is a well defined operator for any $0<\alpha<2m$, especially also at integer points $\frac{\alpha}{2}=p\in \N$. The normalization factor ${\cal C}_{m,n,\alpha}$ of (\ref{FL}) is given by (\ref{basishelpintegral2})

\begin{equation}
 \label{basishelpintegral2p}
A_{m,n,\alpha=2p} ={\cal C}_{m,n,2p}^{-1} = U_{n,2p}V_{m,2p} >0 ,\hspace{2cm} \frac{\alpha}{2} =p \in \N
\end{equation}
where we do not further need to evaluate $V_{m,\alpha=2p}$ here.
For the analysis performed in this paper
the values of the normlization factor (\ref{basishelpintegral2p}) for fractional $\frac{\alpha}{2} \notin \N $ is most relevant:

{\it Explicit determination of  (\ref{Vinteg}) for the fractional case $\frac{\alpha}{2} \notin \N$}

\begin{equation}
 \label{rep1}
\frac{1}{|\xi|^{\alpha +1}} = -\frac{1}{\sin{\frac{\pi\alpha}{2}}}\lim_{\epsilon\rightarrow 0+}\Re(\epsilon +i\xi)^{-\alpha-1} =
\frac{-1}{2\alpha!\sin{\frac{\pi}{2}}\alpha}\int_{-\infty}^{\infty}e^{ik\xi}|k|^{\alpha}{\rm d}k =-\frac{\pi}{\alpha!\sin{\frac{\pi}{2}\alpha}} (-\frac{d^2}{d\xi^2})^{\frac{\alpha}{2}}\delta(\xi) ,\hspace{0.3cm} \frac{\alpha}{2} \notin \N
\end{equation}
where $\sin{\frac{\pi\alpha}{2}} \neq 0$ and we have put $\alpha!=\Gamma(\alpha+1)$.
After some manipulations (\ref{Vinteg}) takes the form

\begin{equation}
 \label{Vmalpha}
V_{m,\alpha}= (-1)^{m+1}\frac{\pi}{2^{\alpha+1}\alpha!\sin{\frac{\pi}{2}\alpha}}\left(D_{\lambda}(1)-D_{\lambda}(-1)\right)^{2m}|
\lambda|^{\alpha}|_{\lambda=0} ,\hspace{1cm} 0<\alpha<2m , \hspace{1cm} \frac{\alpha}{2} \notin \N
\end{equation}
where this relation is undetermined for integer $\frac{\alpha}{2} \in \N$ due to the vanishing of
$\sin{\frac{\pi}{2}\alpha}=0$ in the nominator and the finite difference $\left(D_{\lambda}(1)-D_{\lambda}(-1)\right)^{2m}|
\lambda|^{\alpha}|_{\lambda=0}=0$ in the denominator at these values. However, this relation is well determined in the limiting case (in the sense of L'H\^opital's rule) with $\frac{\alpha}{2} = p\pm \delta $ being infinitesimally close to integer powers $p\in \N_0$ $\delta\rightarrow 0$ (where $\delta$ remains infinitesimally non-zero). 
This gives the possibility to extend (\ref{Vmalpha}) analytically also to the ``forbidden" integer values. However, we emphasize, that there is no need to perform this limiting case if regularization rule (\ref{PV}) is utilized and
taking into account that the finite difference operator in (\ref{Vmalpha}) is cancelling with (\ref{zeroterm}) and hence does not affect the regularized representation of the FL (\ref{universalFL}).

In (\ref{Vmalpha}) the shift operator expression is defined by its binomial series \cite{michel-collet}

\begin{equation}
\label{shiftex}
\begin{array}{l}
(-1)^{m+1}\Delta_{2m}(1)|\lambda|^{\alpha}|_{\lambda=0}= \left(D_{\lambda}(1)-D_{\lambda}(-1)\right)^{2m}|
\lambda|^{\alpha}|_{\lambda=0} \nonumber \\ \nonumber \\
\ds  \left(D_{\lambda}(1)-D_{\lambda}(-1)\right)^{2m}|
\lambda|^{\alpha}|_{\lambda=0} = (-1)^m\sum_{p=-m}^m\frac{2m!}{(m+p)!(m-p)!}(-1)^pD_{\lambda}(2p)|\lambda|^{\alpha}|_{\lambda=0} \nonumber \\ \nonumber \\
\ds \left(D_{\lambda}(1)-D_{\lambda}(-1)\right)^{2m}|
\lambda|^{\alpha}|_{\lambda=0} = 2^{1+\alpha}(-1)^m\sum_{p=1}^m\frac{2m!}{(m+p)!(m-p)!}(-1)^pp^{\alpha}
\end{array}
\end{equation}
where the term for $p=0$ is vanishing since $\alpha>0$ and the symmetry of the terms for $\pm p$ has been used.

The constant (\ref{basishelpintegral}) yields  by using the factorization (\ref{basishelpintegral2})
with (\ref{Vmalpha}) and (\ref{Uint}) the expression

\begin{equation}
 \label{Amnalph}
A_{m,n,\alpha} = {\cal C}_{m,n,\alpha}^{-1}=\frac{\pi^{\frac{n+1}{2}}\Gamma(\frac{\alpha+1}{2})}{2^{\alpha}\Gamma(\frac{\alpha+n}{2})
\Gamma(\alpha+1)\sin{\frac{\alpha\pi}{2}}}
(-1)^{m+1}\left(D_{\lambda}(1)-D_{\lambda}(-1)\right)^{2m}|
\lambda|^{\alpha}|_{\lambda=0} ,\,\,0 <\alpha< 2m , \,\,\, \frac{\alpha}{2}\notin \N
\end{equation}
or with (\ref{shiftex})$_3$

\begin{equation}
 \label{Amnalphexplicit}
\ds A_{m,n,\alpha} = \frac{2\pi^{\frac{n+1}{2}}\Gamma(\frac{\alpha+1}{2})}{\Gamma(\frac{\alpha+n}{2})\Gamma(\alpha+1)
\sin{\frac{\alpha\pi}{2}}} \sum_{p=1}^m \frac{(-1)^{p-1}(2m)!}{(m+p)!(m-p)!}p^{\alpha}
\end{equation}
For $m=1$ (\ref{Amnalph}) takes $A_{m=1,n,\alpha}={\cal C}^{-1}_{m=1,n,\alpha}=2{\cal C}_{n,\alpha}^{-1}$.

\subsection{Explicit determination of the surface integral $U_{n,\alpha}$ of (\ref{Uint}) }
\label{norm-fac}

In this appendix we evaluate the surface integral $U_{n,\alpha}$ defined by the surface integral (\ref{Uint2}) on the $n$-dimensional
unit-ball. For our convenience we first consider the following
integral

\begin{equation}
 \label{gaushelp}
I_{\alpha,n}= \int e^{-r^2} |x|^{\alpha}{\rm d}^n{\bf r}=\int_{|{\hat r}|=1}|{\hat x}|^{\alpha}{\rm d}\Omega({\hat r})\int_0^{\infty}
 e^{-r^2}r^{\alpha+n-1}{\rm d}r=\frac{U_{n,\alpha}}{2}\int_0^{\infty}e^{-\tau}\tau^{\frac{\alpha+n}{2}-1}{\rm d}\tau
=\frac{1}{2}U_{n,\alpha}\Gamma(\frac{\alpha+n}{2})
\end{equation}
where we have put $x/r={\hat x}$.

Let us assume $\alpha\in \R$ and $\alpha >-1$:
(\ref{gaushelp}) is performed over the entire $\R^n$ and we introduce $x=rn_1$ with ${\rm d}^n{\bf r}=r^{n-1}{\rm d}r{\rm d}\Omega({\bf n})$. Then (\ref{gaushelp}) can also be written as

\begin{equation}
 \label{intg1}
I_{\alpha,n}= \int_{-\infty}^{\infty} e^{-x^2} |x|^{\alpha}{\rm d}x \,
\Pi_{j=2}^n \int_{-\infty}^{\infty} e^{-y_j^2}{\rm d}y_j =2\pi^{\frac{n-1}{2}}\int_0^{\infty} e^{-x^2} |x|^{\alpha}{\rm d}x
\end{equation}
and the latter integral assumes the form of a $\Gamma$-function ($\tau=x^2$)

\begin{equation}
\label{gammaalph}
\int_0^{\infty} e^{-x^2} |x|^{\alpha}{\rm d}x = \frac{1}{2}\int_0^{\infty}e^{-\tau}\tau^{\frac{\alpha-1}{2}}{\rm d}\tau=
\frac{1}{2}\left(\frac{\alpha-1}{2}\right)!= \frac{1}{2}\Gamma(\frac{\alpha+1}{2}))
\end{equation}
We observe that (\ref{gammaalph}) is well defined (non-singular) for $\alpha>-1$.
and so (\ref{gaushelp}) assumes

\begin{equation}
\label{ialp}
I_{\alpha,n}=\pi^{\frac{n-1}{2}}\Gamma(\frac{\alpha+1}{2})
\end{equation}
So we obtain with (\ref{gaushelp}) and (\ref{ialp}) for the surface integral (\ref{Uint}) the explicit form

\begin{equation}
 \label{Jalpn}
U_{n,\alpha} = \int_{|{\hat r}|=1}|{\hat x}|^{\alpha}{\rm d}\Omega({\hat r}) = \frac{2I_{\alpha,n}}{\Gamma(\frac{\alpha+n}{2})} =
\frac{2\pi^{\frac{n-1}{2}}\Gamma(\frac{\alpha+1}{2})}{\Gamma(\frac{\alpha+n}{2})}
\end{equation}
which is well defined for $\alpha >-1$.
The dimensions $n=1,2,3$ are relevant, so we have

\begin{equation}
 \label{n1}
U_{n=1,\alpha} = 2 , \hspace{2cm} n=1
\end{equation}

\begin{equation}
 \label{n2}
U_{n=2,\alpha} = \int_0^{2\pi}|\cos{\varphi}|^{\alpha}{\rm d}\varphi=2\sqrt{\pi}\frac{\Gamma(\frac{\alpha+1}{2})}{\Gamma(1+\frac{\alpha}{2})}, \hspace{2cm} n=2
\end{equation}

\begin{equation}
 \label{n3}
U_{n=3,\alpha} =\frac{4\pi}{\alpha+1}  ,\hspace{2cm} n=3
\end{equation}

\end{appendix}

\end{document}